\documentclass[9pt,journal]{IEEEtran}
\newtheorem{theorem}{Theorem}

\newtheorem{lemma}{Lemma}
\newtheorem{definition}{Definition}
\newtheorem{remark}{Remark}

\newtheorem{assumption}{Assumption}

\newtheorem{proposition}{Proposition}
\usepackage{graphicx}
\usepackage{subfigure}
\usepackage{array} 
\usepackage{amsmath}
\usepackage{amssymb}
\usepackage{mathtools}        
\usepackage{latexsym}
\usepackage{mathrsfs}
\usepackage{cite}
\usepackage{enumerate} 
\usepackage{float}
\usepackage{amsfonts} 
\usepackage{epsfig} 
\usepackage{epstopdf}
\usepackage[dvipsnames]{xcolor}
\usepackage{tikz}\newcommand*{\circled}[1]{\lower.7ex\hbox{\tikz\draw (0pt, 0pt)%
		circle (.4em) node {\makebox[1em][c]{\small #1}};}}
\usepackage{color}
\usepackage[scaled]{helvet}
\usepackage{tikz,xcolor,hyperref}
\definecolor{lime}{HTML}{A6CE39}
\DeclareRobustCommand{\orcidicon}{\begin{tikzpicture}\draw[lime, fill=lime] (0,0) circle [radius=0.16] node[white] {{\fontfamily{qag}\selectfont \tiny ID}}; \draw[white, fill=white] (-0.0625,0.095) circle [radius=0.007];  \end{tikzpicture}\hspace{-2mm}}
\foreach \x in {A, ..., Z}{\expandafter\xdef\csname orcid\x\endcsname{\noexpand\href{https://orcid.org/\csname orcidauthor\x\endcsname}{\noexpand\orcidicon}}}

\begin{document}
\title{Prescribed-Time Control and Its Latest Developments}
\author{Yongduan Song$^{*}$\orcidA{},  \IEEEmembership{Fellow,~IEEE},  Hefu Ye\orcidB{}, and Frank L. Lewis\orcidC{},  \IEEEmembership{Fellow,~IEEE}
	\thanks{E-mail address: ydsong@cqu.edu.cn (Y. Song),   yehefu@cqu.edu.cn (H.  Ye), and lewis@uta.edu (F. L. Lewis). }
}
\maketitle

\begin{abstract}  
Prescribed-time (PT) control, originated from \textit{Song et al.}, has gained increasing attention among control community. The salient feature of PT control lies in its ability to achieve system stability within a finite settling time user-assignable in advance irrespective of initial conditions. It is such a unique feature that has enticed many follow-up studies on this technically important area, motivating numerous research advancements. In this article, we provide a comprehensive survey on the recent developments in PT control. Through a concise introduction to the concept of PT control, and a unique taxonomy covering: 1) from robust PT control to adaptive PT control; 2) from PT control for single-input-single-output (SISO) systems to multi-input-multi-output (MIMO) systems; and 3) from PT control for single systems to multi-agent systems, we present an accessible review of this interesting topic. We highlight key techniques, fundamental assumptions adopted in various developments as well as some new design ideas. We also discuss several possibles future research directions towards PT control.
\end{abstract}


 \section{Introduction}\label{sec:introduction}
 \IEEEPARstart{T}he notion of prescribed-time control, originally proposed by \textit{Song, Wang, Holloway and Krstic}\cite{2017-song-prescribed-time}, has brought much vitality to finite-time (FT) control, attracting increasing attention from the control community and motivating numerous follow-up studies on this important field during the past few years [\citen{2019-song-prescribed-RNC}]--[\citen{2022-ye-tac}].  Examining the development history of FT control theory reveals that the related concepts can be traced back to the 1960s, when the concept of FT stability was proposed in \cite{1961-finite-short,1967-weiss-finite,1970-finite-linear}, with application to certain simple linear systems. A system is considered to be FT stable if, given a bounded initial condition, all closed-loop signals are bounded and its states converge to zero (or a residual set) within a specified FT interval.  It is essential to identify FT stability and   asymptotic stability.  In fact, a system can be FT stable but not  asymptotically stable, and vice versa. Asymptotic stability corresponds to the behavior of a system within a sufficiently long (in principle, infinite) time interval, while FT stability is a more practical concept that helps to study the behavior of the closed-loop system over a finite (possibly short) time interval, and hence it finds application whenever it is desired/required that the system states shrink to a certain small threshold (for example, to avoid saturation or the excitation of nonlinear dynamics) within a short period  of time. 
 
 The past few decades have witnessed much progress in FT control of dynamic systems\cite{1998-bhat,1986-Haimo,2020-liuyang-survey,2001-Amato,2003-Levant,2005-Huang,2006-survey-finite,2008-xiaxiaohua-finite-time-lemma,2011-ding-survey,2006-Hong,2018-Yujinpeng-finite-command,2017-wanghonghong-finite-time-lemma}.
 The homogeneous approach, terminal sliding mode control method, and adding a power integrator method are suggested sequentially in an attempt to achieve FT stability for high-order nonlinear systems\cite{2013-Chen-terminal-mode,2005-Bhat,2017sunzongyaofast-finite,2000-LinW,2005-Levant}.
 Although convergence may be pursued in a finite time,  estimation of the settling time relies explicitly on initial conditions. This may limit the application scope of those existing results when little knowledge of plant initial states are accessible. Later on a notion termed fixed-time (FxT) control\cite{2011-polyakov-fixed,2015-polyakov-fixed-time,2018-zuo-overview,2020Polyakov,2014-zuozongyu-fixed-time,2016-zuozongyu-fixed-time,2022-han-TII} has emerged, which employs odd-order  plus fractional-order  feedback to provide various closed-loop system dynamics. The upper bound of the settling time can be estimated without using any information on initial   conditions. 
 
 Despite the benefits of FxT control in the light of settling time estimation, no simple and obvious relationship exists between the control parameters and the intended upper-bound of the settling time. In addition, the settling time under the FxT control is often overestimated,  which may be hundreds or even thousands of times  larger than  the true settling time, resulting in  an inaccurate description of system performance. On the other hand, the settling time is not a directly tuneable parameter for either FT control or FxT control, as it also depends on other controller design parameters.  To alleviate  the problem of overestimation of the settling time while alleviating the dependence of the settling time on design parameters, the predefined-time control (PdT control) approach is exploited in \cite{2018-Sanchez-predefined-time,2021-Predefined,2019-predefined-enhancing,2020-predefined-class}, where the least  upper bound of the settling time can be preset irrespective  of initial conditions and any other design parameters. 
 
 Recently, the classical idea that originated in strategic and tactical missile guidance applications \cite{Zarchan07,1965-HoY-tac-optimal,1973-later-optimal} has been revisited and further applied to high-order nonlinear systems, namely PT control, which inherits the advantages of FT control, FxT control, PdT control and also allows for  presetting the settling time precisely.  This concept is of great importance in many practical engineering applications where transient processes must occur within a given time  (e.g., missile guidance,  multi-agent rendezvous, emergency braking, and  obstacle avoidance in robotic systems, etc.).  
 
 More importantly, the PT control is promising since it is  robust to external disturbances, the control input is always smooth over the transient process, and there is no need for any information on the upper bound of the non-vanishing perturbations in the control design. The key technical design steps for PT control include: converting the original system to a new system by a time-varying transformation (including state scaling, time scaling, and some other technologies), dealing with matching/mismatching uncertainties and unknown control coefficients to construct appropriate Lyapunov inequalities, and selecting the appropriate control gain $k$ to prove the boundedness of all closed-loop signals, especially the boundedness of the control inputs.  Furthermore, because all real PT controllers have infinite gain characteristics as time tends to the pre-set time, they can only be used for a finite time interval. Many infinite time controllers may be integrated with PT algorithms to deliver their infinite time features inside a prescribed-time window, thereby extending the use of PT control systems.  In this article, we perform a complete study on several important theoretical breakthroughs, key technical concerns, and potential research problems in PT control, as well as provide a comprehensive literature survey.

 The study will start next in Section \ref{section2} with an overview of some basic propositions of FT/FxT/PdT and PT control. Section \ref{ptc} lists some specific literature on PT control for SISO systems  and  provides some basic design ideas of prescribed robust and/or adaptive controller design, focused on the introduction of state scaling technology and time scaling technology on PT control. Section \ref{section4} lists some state-of-the-art results on MIMO systems and presents a detailed demonstration of PT control for this type of system. It covers square and non-square  MIMO systems. Section \ref{section5} lists some interesting studies on PT distributed  control and addresses some basic issues of PT control for multi-agent systems. The organization of Sections \ref{section2}-\ref{section5} is shown in Fig. 1. Section \ref{section6} provides some connections between FT  and PT control, and also discusses  some possible open areas of research. 
 \begin{figure} [!htbp]
 	\begin{center}
 		\includegraphics[height=4.3cm]{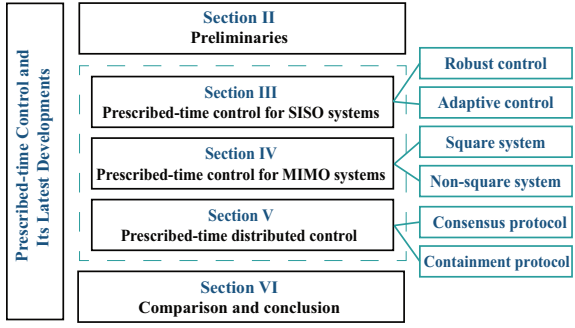}
 		\caption{The organization of this article.}
 	\end{center}\label{fig1}
 \end{figure}
 
 \section{Preliminaries}\label{section2}

 \subsection{Definitions}
 We first consider some basic definitions of infinite-time (asymptotic/exponential) stability:
 \begin{definition}[{[\citen{1991Slotine}], Ch. 4}]\label{definition1}
 	For a non-autonomous system as
 	\begin{equation}\label{autonomous system}
 	\dot x= f(x,t)
 	\end{equation}
 	where $f:\mathbb{R}^n\times [0, \infty)  \rightarrow \mathbb{R}^{n}$ is piecewise continuous in $t$ and locally Lipschitz in $x$. The equilibrium point $x=\mathbf{0}$ is
 	\begin{itemize}
 		\item stable, if there exists a class of $\mathcal{K}$ function $\beta$  such that $
 		\|x(t)\|\leq\beta(\|x(0)\|);$
 		\item asymptotically stable\footnote{If $x(0)\in(-\infty,+\infty)$, here we say that the equilibrium point is global asymptotically stable. 
 		}, if it is stable, and
 		$
 		x(t)\rightarrow\mathbf{0}~  \text{as}  ~t \rightarrow \infty;
 		$ 
 		\item  exponentially stable, if there exist two positive numbers, $\lambda_1$ and $\lambda_2$, such that for sufficiently small $x(0)$, $
 		\|x(t)\|\leq \lambda_1\|x(0)\|e^{-\lambda_2t},~\forall t\geq 0.
 	$	
 	\end{itemize}
 \end{definition}
 The definitions of FT and FxT stability are stated as below:
 \begin{definition}[{[\citen{2020Polyakov}], Ch. 4}]\label{definition2}
 	For a system as (\ref{autonomous system}), the equilibrium point $x =
 	\mathbf{0}$ is 
 	\begin{itemize}
 		\item finite-time stable, if it is stable and there exists a $x(0)$-dependent settling time function $T(x(0))$ such that
 		$
 		x(t)=\mathbf 0  ~ ~ \text{for} ~~t \geq T(x(0));
 	$
 		\item fixed-time stable, if it is stable and the settling time function $T(x(0))$ is upper bounded on $\mathbb{R}$,  i.e., 
 		$
 		\exists ~T_{\max}>0, ~~x(t)=\mathbf 0  ~ ~ \text{for} ~~t \geq T_{\max}.
 	$
 	\end{itemize}
 \end{definition}
Obviously, the terminal time always attaches itself to $x(0)$ in FT control, such  attachment is however removed in FxT control. An astonishing scenario in FT/FxT stability is the PT stability, where the terminal time has nothing to do with initial condition, thus can be user-set freely in advance.

 \subsection{Propositions on Finite-/Fixed-/Predefined-/Prescribed-time Stability}
 Achieving FT stability  for dynamic systems is of special theoretical and practical interest. The typical approach for establishing FT stability is to derive  Lyapunov differential inequalities. Most of these inequalities can be found in the following works which are summarized as a variety of  propositions:
 \begin{proposition}[{[\citen{2005-Bhat}]}]\label{proposition1}
 	For system $\dot x= f(x,t)$, if there exists a $\mathcal{C}^1$ function $V(x)\geq 0$ such that $\dot{V}(x)\leq-k V^{q}(x),$	where $k>0,~0<q<1$, then the closed-loop system  is FT stable and the settling time is calculated by
 	\begin{equation*}
 	\begin{aligned}
 	T:=\frac{1}{k(1-\alpha)}{V^{1-q}(x(0))}.
 	\end{aligned}
 	\end{equation*} 
 \end{proposition}

 \begin{proposition}[{[\citen{2017sunzongyaofast-finite}]}]\label{proposition2}
 	For system $\dot x= f(x,t)$, if there exists a $\mathcal{C}^1$ function $V(x)\geq 0 $ such that  $\dot{V}(x) \leq -k_1 V^{p}(x)-k_2 V^{q}(x),$
 	where $k_1>0,~ k_2>0,~p \geq  1$ and $0<q<1$, then  the closed-loop system is fast FT stable and the settling time is calculated by
 	\begin{equation*}
 	T:=\left\{\begin{array}{l}
 	\frac{1}{k_{2}\left(1-q\right)}+\frac{V^{1-\alpha_{1}}(x(0))-1}{k_{1}\left(1-p\right)}, \quad~~~~~~~ ~p>1 \\
 	\frac{1}{k_{1}\left(1-q\right)} \ln \left(1+\frac{k_{1}}{k_{2}} V^{1-q}(x(0))\right), \quad p=1.
 	\end{array}\right.
 	\end{equation*} 
 \end{proposition}
 
 \begin{proposition}[{[\citen{2008-xiaxiaohua-finite-time-lemma}]}]\label{proposition3}
 	For system $\dot x= f(x,t)$, if there exists a $\mathcal{C}^1$ function $V(x)\geq 0$ such that $\dot{V}(x) \leq -k_1 V^{q}(x)+k_2 V(x)$
 	where $k_1>0,~k_2>0,$ and $ 0<q<1$, then the closed-loop system is  semi-global FT stable and the settling time is calculated by
 	\begin{equation*} 
 	T:=\frac{1}{k_2(1-q)}\ln\left(1-\frac{k_1}{k_2}V^{1-q}(x(0))\right) .
 	\end{equation*}
 \end{proposition}
 \begin{proposition}[{[\citen{2017-wanghonghong-finite-time-lemma}]}]\label{proposition4}
 	For system $\dot x= f(x,t)$, if there exists a $\mathcal{C}^1$ function $V(x)\geq 0 $ such that $	\dot{V}(x) \leq -k V^{q}(x)+\eta$
 	where $k>0, ~0<\eta<\infty$ and $0<q<1$, then the closed-loop system is practical semi-global FT stable and the settling time is calculated by
 	\begin{equation*}
 	T:=\frac{1}{k\theta(1-q)}\left(V^{1-q}(x(0))-\left(\frac{\eta}{k(1-\theta)}\right)^{\frac{1-q}{q}}\right),
 	\end{equation*}	
 	where $0<\theta<1$ is a constant.
 \end{proposition}
 \begin{proposition}[{[\citen{2018-Yujinpeng-finite-command}]}]\label{proposition5}
 	For system $\dot x= f(x,t)$, if there exists a $\mathcal{C}^1$ function $V(x)\geq 0$ such that $
 	\dot{V}(x) \leq -k_1V^{q}(x)-k_2 V(x)+\eta$
 	where $k_1>0, ~k_2>0, ~0<\eta<\infty $ and $0<q<1$, then the closed-loop system is   practical FT stable and the settling time is calculated by 
 	\begin{equation*}
 	T:=\max\left\{\frac{\ln\Big(\frac{k_2\theta V^{1-q}(x(0))+k_1}{k_1}\Big)}{k_2\theta(1-q)},~\frac{\ln\Big(\frac{k_2V^{1-q}(x(0))+\theta k_1}{\theta k_1}\Big)}{k_2(1-q)}\right\}.
 	\end{equation*}
 \end{proposition}
 \begin{proposition}[{[\citen{2011-polyakov-fixed}]}]\label{proposition6}
 	For system $\dot x= f(x,t)$, if there exists a $\mathcal{C}^1$ function $V(x)\geq 0$ such that $
 	\dot{V}(x) \leq -\left(\alpha V^{p}(x)+\beta V^{q}(x)\right)^{k}$
 	where $\alpha>0, ~\beta>0, ~p>0, ~q>0,~ k>0,$ and $ p k<1, ~q k>1$, then  the closed-loop system is FxT stable and the settling time is bounded by
 	\begin{equation*}
 	T:=\frac{1}{\alpha^k(1-pk)}+\frac{1}{\beta^k(qk-1)}.
 	\end{equation*}
 \end{proposition} 
 \begin{proposition}[{[\citen{2020Polyakov}]}]\label{proposition7}
 	For system $\dot x= f(x,t)$, if there exists a $\mathcal{C}^1$ function $V(x)\geq 0 $ such that $
 	\dot{V}(x) \leq -\alpha V^{p}(x)-\beta V^{q}(x)$
 	where $\alpha>0, ~\beta>0,~ p=1-\frac{1}{2 \gamma}, ~q=1+\frac{1}{2 \gamma}, ~\gamma>1$, then the closed-loop system is FxT stable and the settling time is bounded by
 	\begin{equation*}
 	T:=\frac{\pi \gamma}{\sqrt{\alpha\beta}}.
 	\end{equation*}
 \end{proposition} 
 \begin{proposition}[{[\citen{2014-zuozongyu-fixed-time}]}]\label{proposition8}
 	For system $\dot x= f(x,t)$, if there exists a $\mathcal{C}^1$ function $V(x)\geq 0$ such that $
 	\dot{V}(x) \leq -\alpha V^{2-\frac{p}{q}}(x)-\beta V^{\frac{p}{q}}(x)$
 	where $\alpha>0, ~\beta>0, ~q>p>0$ and both $ q $ and $ p $ are odd integers, then   the closed-loop system is FxT stable and the settling time is bounded by
 	\begin{equation*}
 	T:=\frac{q\pi}{2\sqrt{\alpha\beta}(q-p)}.
 	\end{equation*}
 \end{proposition} 
 \begin{proposition}[{[\citen{2016-zuozongyu-fixed-time}]}]\label{proposition9}
 	For system $\dot x= f(x,t)$, if there exists a $\mathcal{C}^1$ function $V(x)\geq 0$ such that $\dot{V}(x) \leq -k_1 V^{\frac{m}{n}}(x)-k_2 V^{\frac{p}{q}}(x)$ 
 	where $k_1>0, ~k_2>0, ~q>p>0, ~m>n>0$ and $ p,~q,~m $ and $ n $ are all odd integers, then  the closed-loop system is FxT stable and the settling time is bounded by
 	\begin{equation*}
 	T:=\frac{1}{k_1}\frac{n}{m-n}+\frac{1}{k_2}\frac{q}{q-p}.
 	\end{equation*}
 \end{proposition} 
 \begin{proposition}[{[\citen{2018-Sanchez-predefined-time}]}]\label{proposition10}
 	For system $\dot x= f(x,t)$, if there exists a $\mathcal{C}^1$ function $V(x)\geq 0$ such that  
 	\begin{equation}\label{predefined}
 	\dot{V}(x) \leq -\frac{1}{pT_p}e^{V^p(x)}V^{1-p}(x)
 	\end{equation}  
 	where $T_p>0, ~0<p\leq 1$,  
 	then the closed-loop system is weakly PdT stable; if the equal sign in (\ref{predefined})  always holds, then  the closed-loop system is strongly PdT stable. The settling time is upper bounded by $T_p$. 
 \end{proposition} 
 \begin{proposition}[{[\citen{2017-song-prescribed-time,2019-song-prescribed-RNC}]}]\label{lemma1}
 	Consider a time-varying function $ \mu(t)=T/(T-t)$, if a $\mathcal{C}^1$  function $ V: [0, T)\rightarrow [0, +\infty)$ satisfies
 	\begin{equation}
 	\label{2.2.2}
 	\dot V(t)  \leq-2 k \mu\left(t\right) V(t)+\frac{\mu\left(t\right)}{4 \theta} d(t)^{2} 
 	\end{equation}
 	for unknown perturbation $ d(t) $ and positive numbers $k,~\theta$, then $ V(t) $ is bounded for $ t\in[0, T). $
 \end{proposition}
 \begin{proposition}\label{lemma1-extend}
 	Consider a   time-varying function  $ \mu(t)=T/(T-t)$, if a  $\mathcal{C}^1$ positive function $ V: [0, T)\rightarrow [0, +\infty)$ satisfies
 	\begin{equation} \label{dV}
 	\dot V(t)  \leq- k \mu\left(t\right) V(t)+|d(t)| 
 	\end{equation}
 	for unknown perturbation $ d(t) $  and a positive number  $k$, then $ V(t) $ is bounded for $ t\in[0, T) $ and $\lim_{t\rightarrow T}V(t)=0$.
 \end{proposition}
 \textit{Proof of Proposition \ref{lemma1-extend}:} Solving the differential inequality (\ref{dV}) gives
 \begin{equation}
 V(t)\leq  e^{-A(t)}V(0)+\sup_{s\in[0,t]}|d(s)|e^{-A(t)}\int_{0}^{t}e^{A(s)} ds
 \end{equation}
 where $A(t)=\int_{0}^{t}k\mu(s) ds$.  Since $\lim_{t\rightarrow T}A(t)=+\infty$ and $\lim_{t\rightarrow T}e^{-A(t)}\int_{0}^{t}e^{A(s)}ds=1/(k\mu)=0$, then $\lim_{t\rightarrow T}V(t)=0$. $\hfill\blacksquare$

The connections and differences among the aforementioned stability notions are conceptually highlighted in Fig. 2. Basically, PT stability is most ``desirable", which covers  PdT stability,  FxT stability, and of course FT stability, and FxT stability implies FT stability, but the reverse does not necessarily hold.
 \begin{figure} [!h]
 	\begin{center}
 		\includegraphics[height=4cm]{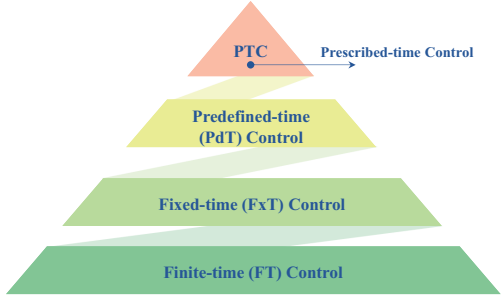}
 		\caption{The relationships among FT/FxT/PdT and PT control. Key features: FT control—the settling time is related to initial conditions and design parameters; FxT control—the settling time can be estimated by an upper bounded function independent of initial conditions; PdT control—the upper bound of the settling time can be user-set freely; and PT control—the exact settling time can be user-assigned arbitrarily.}
 	\end{center}\label{fig2}
 \end{figure}

Propositions \ref{proposition1}-\ref{proposition5} indicate that the terminal time $T$ attaches itself to several  design parameters (e.g., $b$, $c$, $\alpha$, etc.), and the initial system state $V(x(0))$. Propositions \ref{proposition6}-\ref{proposition9} show that the settling time $T$ is bounded by a computable function, which is independent of  the initial condition $V(x(0))$, whereas Propositions \ref{proposition10}-\ref{lemma1-extend} indicate that the settling time $T$ can be pre-set at users' will  irrespective of initial conditions and any other design parameter. To  close this section, we summarize the contents of Propositions \ref{proposition1}-\ref{lemma1-extend} through Table 1. 
 
 \begin{table*}[!htbp] 
 	\centering
 	{\centering Table 1. Related Propositions on FT, FxT, PdT, and PT control (the meaning of the related parameters see Propositions \ref{proposition1}-\ref{lemma1-extend})}
 	\scalebox{1.01}{
 		{\tabcolsep0.05in  		\renewcommand\arraystretch{1.5}
 			\begin{tabular}{|c|c|c|}
 				\hline Proposition & Expression of $\dot V(t)$ & Settling time function $ T $  	\\
 				\hline
 				$1$ & $ \dot{V} {(x)} \leq -k V ^{q}{(x)} $ & $  T:=\frac{1}{k(1-q)}{V^{1-q}(x(0))} $  \\		
 				$2$ & $ \dot{V} {(x)} \leq -k_1 V^{p} {(x)}-k_2 V^{q} {(x)} $ & $T:=\left\{\begin{array}{l}
 				\frac{1}{k_{2}\left(1-q\right)}+\frac{V^{1-p}(x(0))-1}{k_{1}\left(1-p\right)}, \quad~~~~~~~~~~p>1 \\
 				\frac{1}{k_{1}\left(1-q\right)} \ln \left(1+\frac{k_{1}}{k_{2}} V^{1-q}(x(0))\right), \quad p=1
 				\end{array}\right.$  \\	
 				$3$ & $ \dot{V} {(x)} \leq -k_1 V^{\alpha} {(x)}+k_2V {(x)} $ & $ T:=\frac{1}{k_2(1-q)}\ln\left(1-\frac{k_1}{k_2}V^{1-q}(x(0))\right)$ \\
 				\hline
 				$4$ & $ \dot{V} {(x)} \leq -kV ^{q}{(x)}+\eta$ & $  T:=\frac{1}{k\theta(1-q)}\left(V^{1-q}(x(0))-\left(\frac{\eta}{k(1-\theta)}\right)^{\frac{1-q}{q}}\right)$\\
 				$5$ & $ \dot{V} {(x)} \leq-k_1 V^{q} {(x)}-k_2 V {(x)}+\eta $ &   $T:=\max\Big\{\frac{\ln\big(\frac{k_2\theta V^{1-q}(x(0))+k_1}{k_1}\big)}{k_2\theta(1-q)},~\frac{\ln\big(\frac{k_2V^{1-q}(x(0))+\theta k_1}{\theta k_1}\big)}{k_2(1-q)}\Big\}$ 			\\
 				\hline
 				$6$& $ \dot{V} {(x)} \leq -\left(\alpha V^{p} {(x)}+\beta V^{q} {(x)}\right)^{k} $ & $  T:=\frac{1}{\alpha^k(1-pk)}+\frac{1}{\beta^k(qk-1)} $    \\
 				$7$ & $ \dot{V} {(x)} \leq -\alpha V^{p} {(x)}-\beta V^{q} {(x)} $ & $ T:=\frac{\pi \gamma}{\sqrt{\alpha\beta}}  $  \\
 				$8$ & $ \dot{V} {(x)} \leq -\alpha V^{2-\frac{p}{q}} {(x)}-\beta V^{\frac{p}{q}} {(x)} $ & $T:=\frac{q\pi}{2\sqrt{\alpha\beta}(q-p)} $    \\
 				$9$& $ \dot{V} {(x)} \leq -k_1 V^{\frac{m}{n}} {(x)}-k_2 V^{\frac{p}{q}} {(x)} $ & $ T :=\frac{1}{k_1}\frac{n}{m-n}+\frac{1}{k_2}\frac{q}{q-p} $  \\
 				\hline
 				$10$& $
 				\dot{V}(x) \leq -\frac{1}{pT_p}e^{V^p(x)}V^{1-p}(x)$ & $ T :=T_p$  \\	
 				$11$& $
 				\dot{V}(x) \leq -2k\mu(t) V+\frac{\mu(t)}{4\theta}d(t)^2$ & $ T :=T_{\operatorname{user}} $  \\	
 				$12$& $
 				\dot{V}(x) \leq -k\mu(t) V+|d(t)|$ & $ T :=T_{\operatorname{user}} $  \\				
 				\hline
 		\end{tabular}} 
 	}
 \end{table*}

 \section{Prescribed-time control for SISO systems}\label{ptc}
 In this section, we focus on several fundamental topics in PT control, such as robust control and adaptive control based on time-varying feedback, as well as some associated technical concerns including system state convergence, the boundedness of control input, and the boundedness of parameter estimations. Before this, we review the related concepts on PT control.
 
 \subsection{Preliminaries on prescribed-time control}\label{preliminaries}
 \begin{definition}[{[\citen{2019-holloway-prescribed}]}]\label{definition3}
 	System (\ref{autonomous system}) is PT globally uniformly asymptotically stable  in time $T$ if there exist  a function $\mu:\left[0, T\right) \rightarrow \mathbb{R}_{+}$ with $\mu$ increasing to $\infty$ as $t\rightarrow T$ and a class $\mathcal{K} \mathcal{L}$ function $\beta$ such that, 
 	 	\begin{equation}
 	\|x(t)\| \leq \beta\left(\left\|x\left(0\right)\right\|, \mu\left(t \right)  \right), ~~\forall t \in\left[0,  T\right),
 	\end{equation} 	
 	where $T$ is a finite number that can be prescribed in the design.
 \end{definition}

 \begin{definition}[{[\citen{2017-song-prescribed-time}]}]\label{definition4}
 	The system $\dot{x}=f(x, t, d)$ ($d$ represents  non-vanishing perturbations) is  PT globally uniformly asymptotically stable  in time $T$ if there exist class $\mathcal{K L}$ functions $\beta$ and $\beta_{f}$, a class $\mathcal{K}$ function $\gamma$,  and a time-varying function $\mu(t):\left[0, T\right) \rightarrow \mathbb{R}_{+}$ with $\mu(t)$ approaching to $\infty$ as $t\rightarrow T$ such that,  
 	\begin{equation}
 	\|x(t)\| \leq \beta_{f}\left(\beta\left(\|x(0)\|, t\right)+\gamma\big(\|d\|_{\left[0 , t\right]}\big), \mu(t) \right) ,~\forall t \in\left[0,  T\right).
 	\end{equation}
 \end{definition}

 Clearly, Definition \ref{definition4}  is a generalization of Definition \ref{definition3} in the presence of non-vanishing perturbations in the system.

 To achieve PT convergence, two general PT controller design approaches, namely state scaling based approach and time scaling based approach, are used in the literature:
 \begin{itemize}
 	\item State scaling: Using a monotonically increasing  function $\mu(t)$   that grows to infinity in finite time $T$ to scale the state, thereby constructing a new variable $z=\mu x$. A control design that keeps $z$ bounded  will implicitly make $x$ go to zero as $t \rightarrow T$.
 	\item Time scaling: Using a nonlinear temporal transformation $\tau = a(t)$ with $a(t)$ being a
 	function defined such that $a(0) = 0$ and $\lim_{t\rightarrow T}a(t) =\infty.$ Since this time scale transformation maps $t\in[0,T)$  to $\tau\in[0,\infty)$. A control design that achieves asymptotic convergence in the light of the time variable $\tau$ implicitly
 	achieves PT convergence in terms of the time variable $t$.
 \end{itemize}
 The basis of state scaling based design is the monotonically increasing function:
 \begin{equation}\label{state-scaling}
 \mu(t)=\frac{T}{T-t}, ~~~t\in[0,T)
 \end{equation}
 where $T > 0$, with the properties that $\mu(0) =1 $  and $\mu(T)=+\infty.$
 
 In addition, the basis of time scaling based design is a temporal axis mapping $\tau=a(t)$,\footnote{A common time scaling is $t=T(1-e^{-\tau})$, i.e., $a(t)=\ln T -\ln(T-t)$. In this case, $\alpha(\tau)= e^{\tau}/T$.} with the properties defined as follows. Let $a'(t)=\frac{d a}{d t}$ and $\alpha(\tau)=a'(a^{-1}(\tau))$, i.e.,  $\alpha(\tau)$ is the function $\frac{da}{dt}$ expressed in the light of the new time variable $\tau.$ Also, $d\tau=a'(t)dt=\alpha(\tau)dt$.\footnote{
 	Since both $x(t)$ and $\breve{x}(\tau) $ relate to the value of the same signal at the same physical time point represented as $t$ in the original time axis and $\tau$ in the converted time axis, we use the notation $\breve{x}(\tau)$ to express a signal $x(t)$ as a function of the transformed time variable $\tau$, i.e., $x(t)\equiv\breve{x}(\tau)$. Hence, $\dot{\breve{x}}(\tau)=\frac{d\breve{x}}{d\tau}=\frac{dt}{d\tau}\frac{dx}{dt}= \frac{1}{\alpha(\tau)}\frac{dx}{dt}=\frac{1}{\alpha(\tau)}\dot x(t)$.} 
 \begin{itemize}
 	\item $a(0) >0 $  and $a(T)=+\infty$;	
 	\item $a(t)$ is continuously differentiable on $t\in [0, T)$;
 	\item $a'(t)>0$ and grows to infinite as $t\rightarrow T.$
 \end{itemize}
 Here we provide an overview on some typical works in PT control, most of which were initially for stabilization of SISO systems within preset time. Technical issues to be covered include controller structure,  selection of time-varying functions, convergence, robustness and performance, observers,  and output feedback design, etc.:

 The well-known proportional navigation law in tactical and strategic missile guidance  (see, for instance, \cite{Zarchan07,1964-Sidar-ptc,1964-rekasius,1965-HoY-tac-optimal,1973-later-optimal}) provides  the original idea for PT control;   early studies are the state scaling based PT control for nonlinear systems in normal-form\cite{2017-song-prescribed-time,2019-song-prescribed-RNC};   then PT control via time base generator\cite{1995-TGB,2018-Becerra-TGB};
 super-exponential and PT precise tracking control for normal-form systems\cite{2019-wangyujuan-auto-general}; PT observer  \cite{2019-holloway-prescribed-observer} and output feedback design for linear time-invariant (LTI) systems based upon the separation principle\cite{2019-holloway-prescribed}; PT observer for LTI systems with measurement delay\cite{2022-espitia-sensor}; predictor-feedback PT stabilization of LTI systems with input delay\cite{2021-Espitia-TAC-PTC}; PT stabilizing control for LTI systems via a hyperbolic tangent type nonlinear feedback\cite{2022-ye-smc};
PT stabilization of strict-feedback-like systems via a dynamic gain feedback design\cite{2020-NYU-dynamic-high-gain}; time scaling based output feedback design for strict-feedback-like systems\cite{2020-NYU-output,2021-NYU-ACSP}; PT stabilization via adding a power integrator technique\cite{2022-gao-prescribed}; PT estimation and output
 regulation of the linearized schr{\"o}dinger equation\cite{2020-steeves-prescribed}; PT stabilization for stochastic nonlinear systems, where a non-scaling method is used\cite{2021-liwuquan-stochastic-prescribed,2021-Li-prescribed-SSC,2022-Li-prescribed-SSC};  PT control for nonlinear systems within a liner decay rate\cite{2021-shakouri-prescribed-linear-decay}; PT control for normal-form systems, where Fa\`{a} di Bruno’s formula and Bell polynomials are used\cite{2021-Shakouri-prescribed-framework};  frozen-time eigenvalues for prescribed-time-stabilized linear time-varying systems\cite{2022-Shakouri}; PT control via bounded time-varying feedback and parametric Lyapunov equation\cite{2020-zhoubin-prescribed}; parametric Lyapunov equation based output feedback PT control\cite{2021-zhoubin-tac-prescribed}; bounded time-varying feedback based PT control for normal-form systems and satellite
 formation flying\cite{2022-zhoubin-tcs,2022-zhoubin-p-normal};  PT control for $p$-normal nonlinear  systems\cite{2022-zhoubin-prescribed-p}; PT sliding mode control\cite{2021-Tsing-prescribed-sliding,2011-prescribed-time-sliding,
 	2010-harl-reentry};  a general time transformation for PT control\cite{2020-yucelen-time-scaling}; adaptive PT control for strict-feedback systems\cite{2021-TAC-huachanghun-prescribed-adaptive,2022-hua-Cyber}; PT differentiator and switched feedback based PT controller\cite{2021-orlov-switched,2022-orvol-prescribed-differentiator}; PT stabilization of a perturbed chain of integrators within the framework of time-varying homogeneity\cite{2020-chitour-stabilization}; a new stabilization scheme with prescribed settling-time bound are investigated in \cite{2022-orlov-auto} by combining state scaling and time scaling transformations; 
 PTC for affine systems and rigid bodies\cite{2019-peng-specified}; PT and prescribed performance tracking control for certain nonlinear systems\cite{2022-auto-fujun};
 practical PT control, namely the output state/tracking error converges to a certain set within a prescribed  time\cite{2021-zhoubin-practical-ptc-sample,2021-TNNLS-wangjianhui-PPT,2021-siso-practical-prescribed}.
 
 The representative results of the PT control for SISO systems via time-varying feedback are summarized in Table 2. Most of them  are based on state feedback. Due to the difficulties of designing complex uncertain systems,  most results assume that the control coefficients (including the  control direction) of the system model are precisely known without nonvanishing perturbations in the system.  In addition, most results consider only robust control schemes and do not consider adaptive control schemes. Because in adaptive control, it is necessary to guarantee the boundedness of parameter estimation (it seems to be difficult to do this with the state scaling based PT control approach) in addition to the boundedness of the control signal, which usually poses a challenge for the controller design.  The following about robust and adaptive PT control will be addressed.

 \begin{table*}[!htbp] 
 	\centering
 	{\centering Table 2. Technical Differences between Different Prescribed-time Control Literature}\\\vspace{0.2em}
 	\scalebox{0.9}{
 		{\tabcolsep0.05in \renewcommand\arraystretch{1.5}
 			\begin{tabular}{|c|c|c|c|c|c|c|c|c|}
 				\hline $\begin{array}{c}\vspace{-0.5em}\text{Ref.} \end{array}$ 
 				& $\begin{array}{c}\vspace{-0.5em}\text{Output}\\\text{Feedback}\end{array}$ 
 				& $\begin{array}{c}\vspace{-0.5em}\text{Mis-matched}\\\text{Uncertainties}\end{array}$ & $\begin{array}{c}\vspace{-0.5em}\text{Unknown}\\\vspace{-0.5em}\text{Control}\\\text{Cofficient}\end{array}$ 
 				& $\begin{array}{c}\vspace{-0.5em}\text{Nonvanishing}\\\text{Perturbations}\end{array}$
 				&
 				$\begin{array}{c}\vspace{-0.5em}\text{Adaptive with} \\\vspace{-0.5em}\text{Bounded Para-} \\\text{meter Estimation} \end{array}$ 
 				&
 				$\begin{array}{c}\vspace{-0.5em}\text{Input}\\\text{Delay}   \end{array}$ 
 				&
 				$\begin{array}{c}\vspace{-0.5em}\text{Tracking}   \end{array}$ 
 				&
 				$\begin{array}{c}\vspace{-0.5em}\text{State Scaling} ~ \blacksquare\\\vspace{-0.5em}\text{Time Scaling}~\blacktriangle\\\text{Other Method}~\blacklozenge\end{array}$ 
 				\\
 				\hline
 				\cite{2017-song-prescribed-time,2019-song-prescribed-RNC,2019-wangyujuan-auto-general}  & $\times$  & $\times$ & $\surd$ & $\surd$ & $\times$ & $\times$ & $\surd$ & $\blacksquare$ \\		
 				\cite{2019-holloway-prescribed,2019-holloway-prescribed-observer} & $\surd$ & $\times$ & $\times$ & $\times$ & $\times$ & $\times$ & $\times$ & $\blacksquare$ \\
 				\cite{2020-NYU-dynamic-high-gain,2020-NYU-output,2021-NYU-ACSP} & $\surd$ & $\surd$ & $\times$ & $\times$ & $\times$ & $\times$ & $\times$ & $\blacktriangle$\\		
 				\cite{2020-yucelen-time-scaling} & $\times$ & $\times$ & $\times$ & $\times$ & $\times$ & $\times$ & $\surd$ & $\blacktriangle$ \\
 				\cite{2020-zhoubin-prescribed,2021-zhoubin-tac-prescribed,2022-zhoubin-prescribed-p} & $\surd$ & $\surd$ & $\times$ & $\times$ & $\times$ & $\times$ & $\times$ & $\blacklozenge$ \\ 
 				\cite{2021-liwuquan-stochastic-prescribed,2021-Li-prescribed-SSC,2022-Li-prescribed-SSC} & $\times$ & $\surd$ & $\times$ & $\times$ & $\times$ & $\times$ & $\times$ & $\blacklozenge$ \\
 				\textcolor{blue}{\cite{2021-TAC-huachanghun-prescribed-adaptive,2022-hua-Cyber}} & $\times$ & $\surd$ & $\surd$ & $\times$ & $\surd$ & $\times$ & $\times$ & $\blacklozenge$ \\
 				\cite{1995-TGB,2018-Becerra-TGB} & $\times$ & $\times$ & $\times$ & $\times$ & $\times$ & $\times$ & $\surd$ & $\blacklozenge$ \\
 				\cite{2022-espitia-sensor,2021-Espitia-TAC-PTC} & $\surd$ & $\times$ & $\times$ & $\times$ & $\times$ & $\surd$ & $\times$ & $\blacksquare$ \\
 				\hline
 		\end{tabular}}
 	}\\ \vspace{0.3em}
 	\footnotesize {If the method (including the existing  extension of this method) can overcome the limitation, it is marked by $\surd$, otherwise, by $\times$.}
 \end{table*}

 \subsection{Robust prescribed-time control}
 In this subsection, we adopt the state scaling method to design a  control $u(t)$ to stabilize a  scalar system with unknown control coefficient and non-vanishing perturbation in prescribed time $T$. Consider: 
 \begin{equation}\label{dx=bu+f}
 \dot x= b(x,t) u +f(x,t)
 \end{equation}
 where $x$ and $u$ are the state and the control input, respectively, $b(x,t)$   and $f(x,t)$ are nonlinear time-varying functions and satisfy the following assumptions. 
 \begin{assumption}[{[\citen{2017-song-prescribed-time}]}]\label{assumption 1}
 	The function $f(x,t)$ is smooth and satisfies $|f(x,t)|\leq d(t)\psi(x)$, where $d(t)$ is a bounded but unknown perturbation, and $\psi(x)\geq 0$ is a known computable function.
 \end{assumption}
 \begin{assumption}\label{assumption 2}
 	The time-varying function $b(t)$ is away from zero, without losing generality, we assume that $b(t)>0$ and there exists an unknown $\underline{b}$ such that $0<\underline{b} \leq|b(x, t)|<\infty$ for all $x \in \mathbb{R}, t \in [0,+\infty)$.
 \end{assumption}
 \begin{remark}
 	Assumption \ref{assumption 2}  is more general than the one used in \cite{2017-song-prescribed-time}, since the latter requires that $\underline{b}$ be known.  
 \end{remark}
 
 \begin{theorem}\label{theorem1}
 	Under Assumptions 1-2, the closed-loop system consisting of (\ref{dx=bu+f}) and the control law (\ref{u-robust}) is PT stable in the sense of Definition \ref{definition4} and  all internal signals are bounded over  $[0,T)$, 
 	\begin{equation}\label{u-robust}
 	u=-k (\mu x)-\theta (\mu x) \left( \psi+|\dot\mu\mu^{-2}(\mu x)|\right)^2
 	\end{equation}
 	where $k>0$ and $\theta>0$.
 \end{theorem} 
 \textit{Proof:} Denote $\mu x$ by $z$, and denote $ \psi+|\dot\mu\mu^{-2}(\mu x)|$ by $\Phi(x,t)$. Choose a Lyapunov function  as $V=\frac{1}{2\underline{b}}z^2$, then,
 \begin{equation}\label{dV=z^2}
 \begin{aligned}
 \dot V=  &\frac{b}{\underline{b}} \mu z u +\frac{1}{\underline{b}}\mu z \big(\dot\mu \mu^{-2}z+f\big)\\
 \leq & \frac{b}{\underline{b}} \mu z u+ \frac{\max\{1,\|d\|_{[0,t]}\}}{\underline{b}} \mu |z| \Phi (x,t) .
 \end{aligned}
 \end{equation}   
 Let $\Delta={\max\{1,\sup\{|d(t)|\}\}}/{\underline{b}}$, applying Young's inequality with $\theta>0$, we get
 \begin{equation}\label{young-inequality}
 \Delta \mu |z|\Phi(x,t)\leq \theta\mu z^2\Phi^2+\frac{\mu\Delta^2}{4\theta}.
 \end{equation} 
 Note that $b/\underline{b}\geq 1$, then substituting (\ref{u-robust}) and (\ref{young-inequality}) into (\ref{dV=z^2}), we have
 \begin{equation}\label{dV=-kmuV+mud}
 \dot V\leq -k \mu z^2 +\frac{\mu\Delta^2}{4\theta}\leq -2\underline{b}k \mu V+\frac{\mu\Delta^2}{4\theta}.
 \end{equation}
 According to Proposition \ref{lemma1}, (\ref{dV=-kmuV+mud}) results in $V\in\mathcal{L}_{\infty}[0,T)$, and hence $  z\in\mathcal{L}_{\infty}[0,T)$. Furthermore, the state $x=\mu^{-1}\sqrt{2\underline{b}V}$ is bounded and converges to zero as $t\rightarrow T$. Since $\dot\mu\mu^{-2}=1/T$ is bounded, then $\Phi(x,t)$ is bounded, establishing the same for $u(t)$. Therefore, all signals are bounded, and the closed-loop system is PT stable in the sense of Definition \ref{definition4}.
 $\hfill\blacksquare$
 
 \begin{remark}
 	It is worth mentioning that, for system (\ref{dx=bu+f}), as long as $b(x,t)$ satisfies $b(x, t)\geq  \underline{b} > 0$ with  $\underline{b}$ being some unknown constant, the proposed PT control does not need \textit{a priori} information on $b(x,t)$, such simple control algorithm without involving $\underline{b}$ can be readily extended to higher-order systems in normal form\cite{2017-song-prescribed-time}. Other robust PT control results can be found in \cite{2020-NYU-dynamic-high-gain,2020-yucelen-time-scaling,2021-liwuquan-stochastic-prescribed,2022-ye-tac} and the references therein.
 \end{remark}

 \subsection{Adaptive prescribed-time control} 
 It is interesting yet challenging to develop adaptive control schemes to regulate the system state to zero in a prescribed time. So far, the related results in this area are very limited.  The following subsections present  three basic frameworks of adaptive PT control through a scalar system: (Section \ref{c.1}) adaptive design for systems with time-invariant parameters; (Section \ref{c.2}) adaptive design for systems with time-varying parameters; and (Section \ref{c.3}) adaptive Nussbaum gain design for systems with time-varying parameters.
 
 We use time scaling method to develop our adaptive control design and hence it is necessary to restate some basic concepts: $i)$ $\breve{x}(\tau)=x(t)$; $ii)$ $\alpha(\tau)\dot{\breve{x}}(\tau)=\dot{x}(t)$; and $iii)$ $\alpha(\tau)>0$.
 
 \subsubsection{Design for systems with time-invariant parameters}\label{c.1} 
 \begin{assumption}[{[\citen{2021-TAC-huachanghun-prescribed-adaptive}]}]\label{assumption3}
 	The nonlinearity  $f(x,t)$  can be parameterized as   $f(x,t)=\theta \psi(x)$ with $\psi(x)$ being a known smooth function and $\psi(0)=0$, and $\theta$ being an unknown constant.
 \end{assumption}
 \begin{assumption}[{[\citen{2008-WenChangyun}]}]\label{assumption4}
 	The function $b(x,t)$, called control coefficient, satisfies $b(x,t)\equiv b$ with $b $ being an unknown nonzero constant.  The sign of $b$ is available  for control design. Furthermore, we assume that there exists a known constant $\underline{b}$ satisfies $\underline{b}<b$. 
 \end{assumption}
 
 Since $\psi\in\mathcal{C}^1$ and $\psi(0)=0$, then by Hadamard's Lemma, we know that there exists a known smooth mapping $\bar\psi$ such that $\psi=\bar{\psi}x$.
 \begin{theorem}\label{theorem 2}
 	Under Assumptions \ref{assumption3}-\ref{assumption4}, the closed-loop system consisting of (\ref{dx=bu+f}) and the adaptive control law (\ref{u-adaptive}) is PT stable in the sense of Definition \ref{definition3} and all internal signals are bounded over  $[0,T)$, 
 	\begin{equation}\label{u-adaptive}
 	\left\{\begin{array}{l}
 	u(t)=\hat{\rho}(t)\bar{u}(t), \\
 	\left\{\begin{array}{l}
 	\bar{u}(t)=-\Big(ka'(t)x(t)+ \frac{1}{2}\hat{\theta}^2x(t) +\frac{1}{2}\bar\psi^2 x(t)\Big),\\
 	\dot{\hat{\theta}}(t)=\gamma_{\theta}x(t)\psi(x),~~~~~~~~~~~~~~~\hat\theta(0)\geq 0,\\
 	\dot{\hat{\rho}}(t)=-\gamma_{\rho}\operatorname{sgn}(b)x(t)\bar{u}(t) 
 	\end{array}\right.
 	\end{array}\right.
 	\end{equation}	 
 	where $k> 1/{({\hat{\rho}(0)\underline{b}})},$  $\gamma_{\theta}>0$, $\gamma_{\rho}>0$ and $a'(t)=\frac{da}{dt}$ is a time-varying function as defined in footnote 2, the initial value of $\rho$ is chosen as $\rho(0)>0$ for $b>0$  (or $\rho(0)<0$ for  $b<0$).
 \end{theorem}
 \textit{Proof:} According to footnote 3, we know that $\breve{x}(\tau)=x(t)$ and $\alpha(\tau)\dot{\breve{x}}(\tau)=\dot{x}(t)$. By using Assumption \ref{assumption3}, we rewrite (\ref{dx=bu+f}) as
 \begin{equation}\label{dbrevex}
 \dot{\breve{x}}=\frac{1}{\alpha(\tau)}(bu+\theta\psi)=\frac{1}{\alpha(\tau)}\left(b \breve{u}(\tau)+\theta\breve{\psi}(\breve{x})\right).
 \end{equation}
 Let $\breve{u}=\hat{\rho}(\tau)\bar{u}(\tau)$, and choose a Lyapunov function $V(\tau):[0,+\infty)\rightarrow [0,+\infty)$ as
 \begin{equation}\label{V-tau-theorem2}
 V(\tau)=\frac{1}{2}\breve{x}^2+\frac{1}{2\gamma_{\theta}}\left(\theta-\hat{\theta}(\tau)\right)^2+\frac{|b|}{2\gamma_{\rho}}\left(\frac{1}{b}-\hat{\rho}(\tau)\right)^2.
 \end{equation}
 The derivative of $V(\tau)$ along the trajectory of the system (\ref{dbrevex})  is
 \begin{equation}\label{V-tau}
 \begin{aligned}
 \frac{dV(\tau)}{d\tau}=&\frac{\breve{x}}{\alpha(\tau)}\left(\bar{u}(\tau)+\hat\theta(\tau) \breve{\psi}\right)+\frac{(\theta-\hat{\theta})}{\gamma_{\theta}}\left(\frac{\gamma_{\theta}\breve{x}\breve{\psi}}{\alpha(\tau)}-\frac{d\hat{\theta}}{d\tau}\right)\\
 &-\frac{|b|}{\gamma_{\rho}}\left(\frac{1}{b}-\hat{\rho}\right)\left(\frac{\gamma_{\rho}\operatorname{sgn}(b)}{\alpha(\tau)}\breve{x}\bar{u}(\tau)+\frac{d\hat\rho}{d\tau}\right).
 \end{aligned}
 \end{equation}
 Note that the control law and update laws designed in Theorem \ref{theorem 2} are equivalent to
 \begin{equation}\label{u-tau}
 \begin{aligned}
 &\bar{u}(\tau)=-k\alpha{(\tau)}\breve{x}-\frac{1}{2}\hat\theta^2{(\tau)}\breve{x}-\frac{1}{2}\breve{\bar\psi}^2\breve{x}\\
 & \frac{d\hat\theta(\tau)}{d\tau}=\frac{1}{\alpha{(\tau)}}\gamma_{\theta}\breve{x}\breve{\psi},~~\frac{d\hat\rho(\tau)}{d\tau}=-\frac{1}{\alpha{(\tau)}}\gamma_{\rho}\operatorname{sgn}(b)\breve{x}\bar{u}(\tau).
 \end{aligned}
 \end{equation}
 Since $-\frac{1}{2}\hat\theta^2{(\tau)}\breve{x}^2-\frac{1}{2}\breve{\bar\psi}^2\breve{x}^2+\hat\theta(\tau) \breve{\psi}\breve{x}\leq 0$, then by
 substituting (\ref{u-tau}) into (\ref{V-tau}), we get
 \begin{equation}\label{xx}
 \frac{dV(\tau)}{d\tau}=-k\breve{x}^2(\tau)\leq 0.
 \end{equation}
 It follows that from (\ref{xx}) that $V(\tau)\in \mathcal{L}_\infty$, which indicates that $\breve{x}(\tau)\in \mathcal{L}_2\cap\mathcal{L}_{\infty}$.  In view of $\dot{\breve{x}}\in\mathcal{L}_{\infty}$, it follows from  Barbalat's Lemma that $\lim_{\tau\rightarrow\infty}\breve{x}(\tau)=0$ (i.e., $\lim_{t\rightarrow T}x(t)=0$).  Furthermore, according to (\ref{u-adaptive}) and Assumption \ref{assumption3} we know  that  there exists a number $L$ such that $|d{\hat\theta}(\tau)/d\tau|\leq  L \breve{x}^2$, and hence $d{\hat\theta}(\tau)/d\tau\in\mathcal{L}_1$. In view of the argument of Theorem 3.1 in \cite{Krstic1996}, it follows that $\hat\theta(\tau)$ has a limit as $\tau\rightarrow\infty$. Similarly, one can conclude that  $\hat\rho(\tau)$ has a limit as $\tau\rightarrow\infty$. In addition, it follows from (\ref{u-adaptive}) that\footnote{It is important to ensure that $x\bar u \leq 0$ since  this guarantees the monotonicity of $\hat{\rho}(t)$   and thus allows to explicitly pick a suitable control gain $k$.}  $\dot{\hat{\rho}}(t)=\gamma_{\rho}\operatorname{sgn}(b)x\bar u=\gamma_{\rho}\operatorname{sgn}(b)(ka'(t)+\hat{\theta}^2/2+\bar{\psi}^2/2)x^2\geq 0$ for $\forall b>0$ and $\dot{\hat{\rho}}(t)\leq 0$ for $\forall b<0$. Therefore, one can conclude that $\hat{\rho}\geq0$ if we choose $\hat{\rho}(0)>0$ when $b>0$, and that $\hat{\rho}\leq0$ if we choose $\hat{\rho}(0)<0$ when $b<0$, which also yields $\hat\rho(t)b>0$.
 Such conclusion is useful for  choosing a suitable control gain $k$, as seen shortly.
 
 To proceed, we rewrite the closed-loop dynamics as 
 \begin{equation}\label{L-dx}
 \dot x=-kb\hat{\rho}a'(t)x- \frac{1}{2}b\hat{\rho}(\hat{\theta}^2x+\bar{\psi}^2x) + \theta \psi
 \end{equation}
 Recall footnote 2, we know that $a(t)=\ln(T/(T-t))$, and
 $$a(0)=1,~~a(T)=+\infty,~~a'(t)=\frac{1}{T-t}.$$
 Then, (\ref{L-dx}) can be simplified as
 \begin{equation}\label{ODE}
 \dot{x}=-\frac{k(t)}{T-t}x+f(x,t)
 \end{equation} 
 where $k(t)=kb\hat{\rho}(t)$ is a positive function and $f(x,t)=- \frac{1}{2}b\hat{\rho}(\hat{\theta}^2x+\bar{\psi}^2x) + \theta \psi$ is a bounded function. Solving the differential inequality (\ref{ODE})   gives:
 \begin{equation} \label{x(t)}
 \begin{aligned}
 x(t)=e^{A(t)} \int_0^t \frac{f(x,s)}{e^{A(s)}}ds+e^{A(t)} x(0), ~A(t)= \int_0^t-\frac{k(s)}{T-s}ds.
 \end{aligned}
 \end{equation} 
 To show the boundedness of $u(t)$, we  state the following Lemma:

 \begin{lemma}\label{Lemma1}
 	For (\ref{x(t)}), if $\lim_{t\rightarrow T}f(x,t)=0$ and a constant $k_{\min}=\inf\{k(t)\}$ satisfies $k_{\min}>1$, then the following equations hold
 	$$\lim_{t\rightarrow T}\frac{e^{A(t)}}{T-t}=0,~~\lim_{t\rightarrow T}\frac{x}{T-t}=0.$$ 
 \end{lemma} 
 \textit{Proof:} It is straightforward  to prove that $$\lim_{t\rightarrow T}\frac{e^{A(t)}}{T-t}=\lim_{t\rightarrow T}\frac{e^{-\int_0^T\frac{k(t)}{T-t}dt}}{T-t}\geq 0,$$ and
 \begin{equation*}  
 \begin{aligned}
 \lim_{t\rightarrow T}\frac{e^{-\int_0^T\frac{k(t)}{T-t}dt}}{T-t}&\leq\lim_{t\rightarrow T}\frac{e^{-\int_0^T\frac{k_{\min}}{T-t}dt}}{T-t}=\lim_{t\rightarrow T}\frac{e^{k_{\min}\ln(T-t)}}{T-t}\\
 &=\lim_{t\rightarrow T}(T-t)^{k_{\min}-1}=0.
 \end{aligned}
 \end{equation*}
 According to Squeeze Theorem, we obtain $\lim_{t\rightarrow T}\frac{e^{A(t)}}{T-t}=0$. Next, we continue to prove $\lim_{t\rightarrow T}\frac{x}{T-t}=0.$ Dividing both sides of   (\ref{x(t)}) by $(T - t)$, we have
 \begin{equation}\label{x/t} 
 \begin{aligned}
 \frac{x}{T-t}=\frac{e^{A(t)}}{T-t} \int_0^t \frac{f(x,s)}{e^{A(s)}}ds+\frac{e^{A(t)}}{T-t} x(0)
 \end{aligned}
 \end{equation}
 As $t\rightarrow T$,   the last term on the right-hand side of (\ref{x/t}) converges to zero since $\lim_{t\rightarrow T} {e^{A(t)}}/({T-t})=0$ and $x(0)$ is bounded. Applying L'H$\hat{\text{o}}$pital's Rule to the first term on the right-hand side of (\ref{x/t}), we have 
 \begin{equation} 
 \begin{aligned}
 &\lim_{t\rightarrow T}\frac{e^{A(t)}}{T-t}\int_0^t \frac{f(x,s)}{e^{A(s)}}ds=\lim_{t\rightarrow T}\frac{\int_0^t {e^{-A(s)}}{f(x,s)}ds}{e^{-A(t)}(T-t)}\\
 &=\lim_{t\rightarrow T}\frac{  {e^{-A(t)}} {f(x,t)} }{\left(\frac{k(t)}{T-t}\right)e^{-A(t)}(T-t)-e^{-A(t)}} =\lim_{t\rightarrow T} \frac{f(x,t)}{k(t)-1}.
 \end{aligned}
 \end{equation}
 Since $k(t)\geq k_{\min}>1$ and $\lim_{t\rightarrow T}f(x,t)=0$, then $\lim_{t\rightarrow T}\frac{e^{A(t)}}{T-t}\int_0^t \frac{f(x,s)}{e^{A(s)}}ds=0$, implying $\lim_{t\rightarrow T}\frac{x}{T-t}=0.$ The proof of Lemma \ref{Lemma1} is completed. 
 
 According to Theorem 1, we know that $$u(t)=-\hat\rho(t)\left( \frac{k x(t)}{T-t}+ \frac{1}{2}\hat{\theta}^2x(t) +\frac{1}{2}\bar\psi^2 x(t)\right).$$
 In terms of Lemma \ref{Lemma1}, we obtain that the control input $u(t)$ is bounded over $[0,T)$ and   $\lim_{t\rightarrow T}u(t)=0$. Therefore,  the closed-loop system is PT stable in the sense of Definition 3. 
 $\hfill\blacktriangle$
 \begin{remark}
 	Since $|\hat\rho(t)|$ is  a monotone increasing function and $b\hat{\rho}>0$, then $k_{\min}=\hat{\rho}(0)\underline{b}$. Therefore,  we only need to pick $k>1/{({\hat{\rho}(0)\underline{b}})}$ to ensure that $k_{\min}>1$. Particularly, when the control coefficient $b$ is known, as assumed in \cite{2021-TAC-huachanghun-prescribed-adaptive}, there is  no need for using Lemma \ref{Lemma1}, we just need to choose $k>1$. 
 \end{remark}

 \subsubsection{Design for systems with time-varying parameters}\label{c.2}
 \begin{assumption}\label{assumption5}
 	The nonlinearity $f(x,t)$  satisfies that $f(x,t)=\theta(t) \psi(x)$, where   $\psi(x)$ is a known smooth function, $\psi(0)=0$, and the time-varying parameter $\theta(t)$ takes values in an unknown compact set, i.e., there exists an unknown constant $\delta_{\theta}$ such that $|\theta(t)|\leq \delta_{\theta}$.
 \end{assumption}
 \begin{remark}
 	Such Assumption is more general than the one used in \cite{2021-TAC-huachanghun-prescribed-adaptive}, since the latter requires that $\theta(t)$ be time-invariant. It is also more general than the Assumption used in \cite{2020-chenkaiwen-adaptive} because the latter requires that $\delta_{{\theta}}$ be known. 
 \end{remark}
 \begin{theorem}\label{theorem 3}
 	Under Assumptions \ref{assumption4}-\ref{assumption5}, the closed-loop system  consisting of (\ref{dx=bu+f}) and the adaptive control law (\ref{u-adaptive-time-varying}) is PT stable in the sense of Definition \ref{definition3} and  all internal signals are bounded over $[0,T)$, 
 		\begin{equation}\label{u-adaptive-time-varying}
 	 \left\{\begin{array}{l}
 	u(t)=\hat{\rho}(t)\bar{u}(t), \\
 	\left\{\begin{array}{l}
 \bar{u}(t)=-\Big(ka'(t)x(t)+\frac{1}{2}\hat{\theta}^2(t)x(t) +\frac{1}{2}\bar\psi^2 x(t)+v(t)\Big),\\
 v(t)=\frac{\hat{\delta}_{\Delta_{\theta}}}{2} {x}(t)(1+\bar\psi^2),\\
 \dot{\hat{\delta}}_{\Delta_{\theta}}(t)=\frac{\gamma_{\delta}}{2}x^2(t)(1+\bar\psi^2),~~~~~\hat\delta_{\Delta_{\theta}}(0)\geq 0\\
 \dot{\hat{\theta}}(t)=\gamma_{\theta}x(t)\psi(x),~~~~~~~~~~~~~~~\hat\theta(0)\geq 0\\
 \dot{\hat{\rho}}(t)=-\gamma_{\rho}\operatorname{sgn}(b)x(t)\bar{u}(t), 
    \end{array}\right.
 	\end{array}\right.
 	\end{equation}	 	
 	where $k>1/{({\hat{\rho}(0)\underline{b}})}$, $\gamma_{\delta}>0$, $\gamma_{\theta}>0$, $\gamma_{\rho}>0$ and $a'(t)=\frac{da}{dt}$ is a time-varying function as defined in Section \ref{preliminaries}.  The initial value of $\rho$ is chosen as $\rho(0)>0$ for $b>0$  (or $\rho(0)<0$ for  $b<0$).
 \end{theorem}
 \textit{Proof:} Similar to the proof of Theorem \ref{theorem 2}, 
 we first rewrite (\ref{dx=bu+f}) as 
 \begin{equation}
 \dot{\breve{x}}=\frac{1}{\alpha(\tau)}\left(b \breve{u}(\tau)+\hat\theta\breve{\psi}+ (\ell_{\theta}-\hat{\theta} )\breve{\psi}+\Delta_{\theta}\breve{\psi}\right)  
 \end{equation}
 with $\ell_{\theta}$ being some constant and $\Delta_{\theta}=\theta(\tau)-\ell_{\theta}$. We then choose a Lyapunov function $V(\tau):[0,+\infty)\rightarrow [0,+\infty)$ as
 \begin{equation}
 \begin{aligned}
 V(\tau)=&\frac{1}{2}\breve{x}^2+\frac{1}{2\gamma_{\theta}}\left(\ell_\theta-\hat{\theta}(\tau)\right)^2+\frac{|b|}{2\gamma_{\rho}}\left(\frac{1}{b}-\hat{\rho}(\tau)\right)^2\\
 &+\frac{1}{2\gamma_{\delta}}\left(\delta_{\Delta_{\theta}}-\hat{\delta}_{\Delta_{\theta}}\right)^2.
 \end{aligned}
 \end{equation}
 With (\ref{u-adaptive-time-varying}), it follows that
 \begin{equation}
 \begin{aligned}
 \frac{dV(\tau)}{d\tau}\leq&-k\breve{x}^2(\tau)-\frac{1}{\alpha(\tau)}\left(\breve{x} v(\tau)-\breve{x}\Delta_{\theta}\breve{\psi}\right)\\
 &-\frac{1}{\alpha(\tau)}\left(\delta_{\Delta_{\theta}}-\hat{\delta}_{\Delta_{\theta}}\right)\breve{x}^2(1+\bar{\psi}^2(\tau)).
 \end{aligned}
 \end{equation}
 Since  
 \begin{equation}\label{Young's}
 \breve{x}\Delta_{\theta}\breve{\psi}=\Delta_{\theta}\bar{\psi}(\tau)\breve{x}^2\leq \frac{\delta_{\Delta_{\theta}}}{2}\breve{x}^2\bar{\psi}^2(\tau)+\frac{\delta_{\Delta_{\theta}}}{2}\breve{x}^2,
 \end{equation} 
 then substituting  $v$ into (\ref{Young's}) yields $\frac{dV}{d\tau}\leq -k\breve{x}^2\leq 0$.  Thus, according to an analysis similar to that in the proof of Theorem \ref{theorem 2}, it can be concluded that all signals are bounded and   the closed-loop system is PT stable in the sense of Definition \ref{definition3}.
 $\hfill\blacktriangle$

 \subsubsection{Adaptive Nussbaum gain design}\label{c.3}
 \begin{assumption}\label{assumption6}
 	The function $b(x,t)$, called control coefficient, is away from zero and takes values in a compact set. However, its magnitude and sign are unknown. There exists a known positive constant $\underline{b}$ satisfies $\underline{b}\leq |b(x,t)|$. 
 \end{assumption} 
 \begin{lemma}[{[\citen{2019-chenzhiyongauto}]}]\label{lemma2}
 	Consider two $\mathcal{C}^{\infty}$ positive functions $ {V}(t)$ : $[0, \infty) \mapsto \mathbb{R}^{+}$ and $ \mathcal{N}(t): [0, \infty) \mapsto \mathbb{R} ^{+}$. Let $b(t):[0, \infty) \mapsto[\underline{b}, \bar{b}]$ for two constants $\underline{b}$ and $\bar{b}$ satisfying $\underline{b} \bar{b}>0$. If, for $\forall t \geq 0$,
 	\begin{equation}
 	\begin{aligned}
 	&\dot{V}(t) \leq(b(t) \mathcal{N}(\xi)+1) \dot{\xi}(t)  ,~~\dot{\xi}(t) \geq 0,  
 	\end{aligned}
 	\end{equation}
 	for an enhanced Nussbaum function  $\mathcal{N}$, then  $\xi(t)$ and $V(t)$ are bounded over the whole time interval $[0, \infty)$.
 \end{lemma}
 \begin{theorem}\label{theorem 4}
 	Under Assumptions \ref{assumption5}-\ref{assumption6},  all internal signals are bounded over  $[0,T)$ and system state $x(t)$ converges to a compact set within preset time $T$, if the control law and update laws are designed as  
 		\begin{equation}\label{u-adaptive-nussbaum}
 	\left\{\begin{array}{l}
 	u(t)=\mathcal{N}(\xi)\bar{u}(t),\\
 	\left\{\begin{array}{l}
 	\bar{u}(t)=ka'(t)x(t)+\frac{1}{2}\left(1+\hat{\theta}^2\bar\psi^2+\hat{\delta}_{\Delta_{\theta}}(1+\bar\psi^2)\right)x(t),\\
 \dot{\hat{\delta}}_{\Delta_{\theta}}(t)=\gamma_{\delta}x^2(t)(1+\bar\psi^2),~~~~~~\hat\delta_{\Delta_{\theta}}(0)\geq 0\\
 	\dot{\hat{\theta}}(t)=\gamma_{\theta}x(t)\psi(x),~~~~~~~~~~~~~~~\hat\theta(0)\geq 0\\
 	\dot{\xi}(t)=x(t)\bar u(t),~~~~~~~~~ ~~~~~~~~~~\xi(0)>0
 	\end{array}\right.
 	\end{array}\right.
 	\end{equation}
 	where $k>0,$  $\gamma_{\delta}>0$, $\gamma_{\theta}>0$, $a'(t)=\frac{da}{dt}$ is a time-varying function as defined in Section \ref{preliminaries} and $\mathcal{N}(\xi)$ is an enhanced type \textit{B-L} Nussbaum function as defined in [\citen{2019-chenzhiyongauto}, Definition 4.2].
 \end{theorem}
 \textit{Proof:} Firstly, we  rewrite (\ref{dx=bu+f}) as 
 \begin{equation}\label{xxx}
 \dot{\breve{x}}=\frac{1}{\alpha(\tau)}\left(\breve{b}(\tau)\mathcal{\breve{N}}(\xi) \breve{u}(\tau)+\hat\theta\breve{\psi}+ (\ell_{\theta}-\hat{\theta} )\breve{\psi}+\Delta_{\theta}\breve{\psi}\right)  
 \end{equation}
 with $\ell_{\theta}$ being an unknown constant and $\Delta_{\theta}=\theta(\tau)-\ell_{\theta}$. Then, choosing a Lyapunov function $V(\tau)$ candidate as
 \begin{equation}
 \begin{aligned}
 V(\tau)=&\frac{1}{2}\breve{x}^2+\frac{1}{2\gamma_{\theta}}\left(\ell_\theta-\hat{\theta}(\tau)\right)^2+\frac{1}{2\gamma_{\delta}}\left(\delta_{\Delta_{\theta}}-\hat{\delta}_{\Delta_{\theta}}\right)^2.
 \end{aligned}
 \end{equation}
 Taking derivative of $V(\tau)$ along the trajectory of (\ref{xxx}), we get 
 \begin{equation}\label{dv-nussbaum-1}
 \begin{aligned}
 \frac{dV(\tau)}{d\tau}=&\left(\breve{b}(\tau)\mathcal{\breve{N}}(\xi)+1\right)\frac{d\xi}{d\tau}-\frac{1}{\alpha(\tau)}\breve{x}\bar{u}+\frac{1}{\alpha(\tau)}\hat{\theta}\breve{\psi}\\
 &+\frac{1}{\alpha(\tau)}\breve{x}\Delta_{\theta}\breve{\psi}+\frac{1}{\gamma_{\theta}}(\ell_{\theta}-\hat{\theta})\left(\frac{1}{\alpha(\tau)}\gamma_{\theta}\breve{x}\breve{\psi}-\frac{d\theta}{d\tau}\right)\\
 &+\frac{1}{ \gamma_{\delta}}\left(\delta_{\Delta_{\theta}}-\hat{\delta}_{\Delta_{\theta}}\right)\Big(- \frac{d{\hat\delta}_{\Delta_{\theta}}}{d\tau}\Big) .
 \end{aligned}
 \end{equation}
 Inserting the control law designed in (\ref{u-adaptive-nussbaum})  into (\ref{dv-nussbaum-1}), yields
 \begin{equation}\label{dv-nussbaum-3}
 \frac{dV(\tau)}{d\tau}\leq \left(\breve{b}(\tau)\mathcal{\breve{N}}(\xi)+1\right)\frac{d\xi}{d\tau},
 \end{equation} 
 where $\frac{d\xi}{d\tau}=\breve{x}\bar{u}(\tau)=k\alpha(\tau)\breve{x}^2(\tau)+\frac{1}{2}\left(1+\hat{\theta}^2\bar\psi^2\right)\breve{x}^2(\tau)+\frac{1}{2}\Big(\hat{\delta}_{\Delta_{\theta}}(1+\bar\psi^2)\Big)\breve{x}^2(\tau)\geq 0, ~\forall \tau\geq0$. Thus,  it follows from Lemma \ref{lemma2} that  $V(\tau)\in\mathcal{L}_{\infty}$ and $\xi\in\mathcal{L}_{\infty}$. Note that the boundedness of $\mathcal{N}(\xi)$ is guaranteed by the boundedness of $\xi$. Therefore, it can be concluded that $\dot{\breve x}\in\mathcal{L}_{\infty}$, which further indicates that $\lim_{\tau\rightarrow\infty}\breve{x}(\tau)=0$ via Barbalat's Lemma. In addition, in terms of the analysis similar to that in the proof of Theorem \ref{theorem 2}, the boundedness of all closed-loop signals can be guaranteed and hence  the closed-loop system is PT stable in the sense of Definition \ref{definition3}.
 $\hfill\blacktriangle$
 
 \begin{remark}
 	It is noted that the adaptive PT control is developed for the system with unknown yet time-varying parameters in both feedback and input channels. These parameters are not slowly time-varying, but rather, are fast time-varying or even involve abrupt changes, thereby making the controller design quite challenging. Although the control algorithm is based on the first-order system, the fundamental idea and the key design steps are worth extending to more general systems. In addition, since neural network (NN) can be combined with robust adaptive to deal with modeling uncertainties, how to compensate the NN reconstruction error to get PT stability represents an increasing topic for future study.
 \end{remark}

 \section{Prescribed-time control for mimo systems}\label{section4}
PT control for MIMO nonlinear systems is an open area of research that is both theoretically and practically important and urgent, especially with new problems arising from emerging applications such as missile guidance,  accurate and timely weather forecasting, aircraft and spacecraft flight control,  and obstacle avoidance in  robotic systems, all of which require new control technologies for time optimization. There have been few research on PT control for MIMO nonlinear systems, particularly when the control gain matrix is unknown, and essentially no findings that can provide PT stabilization, regulation, or tracking.  In \cite{2018-Becerra-TGB}, by using time based generators, a PT control algorithm is  applied to a 7-DoF robot manipulator with a precondition  that all information in the control gain matrix is available. In \cite{2020-zhoubin-prescribed}, a parametric Lyapunov function based PT controller is applied to a  spacecraft rendezvous control system, where the mathematical model of such system can be viewed as a MIMO linear system. In \cite{2021-yilang-EL}, a PT regulation method is developed for the Euler-Lagrange system with known inertia matrix.  In \cite{2022-ye-tac}, PT tracking control for MIMO systems with unknown control gain matrix and non-vanishing uncertainties are studied.
 In addition, some other studies consider  the practical PT tracking control (see, for instance, \cite{2021-cao-cyber-practical}), whose basic idea is to introduce a smooth function that can converge to a given value at the prescribed time, and to convert the original constrained system into an unconstrained one by using the idea of coordinate transformation similar to that in the prescribed performance control theory \cite{2008-bechlioulis-robust}, and finally to obtain the tracking error of the original system that can converge to a given accuracy at the prescribed time by proving the boundedness of the converted system. 
 In the following sections, we  introduce a powerful design approach for MIMO system  that applies not only to square systems but also to non-square systems.     
 
 \subsection{Square system}
 Consider a MIMO nonlinear system  as follows:
 \begin{equation}\label{dX=BU+F}
 \dot X=\mathbf{B}(X,t)U+F(X,t)
 \end{equation}
 where  $U\in\mathbb{R}^n$ and $X\in \mathbb{R}^n$  are the input and the state vector, respectively. $F(X,t)=[f_1,\cdots,f_n]^{\top}\in\mathbb{R}^n$ denotes the modeling uncertainties and external perturbations and each $f_i$ satisfies Assumption \ref{assumption 1}, i.e., $\|F\|\leq d(t) \Psi $ with $\Psi=[\psi_1,\cdots,\psi_n]^{\top}\in\mathbb{R}^n.$  
 \begin{assumption}[{[\citen{2022-ye-tac}]}]\label{assumption7}
 	The matrix $\mathbf{B}(X,t)\in\mathbb{R}^{n\times n}$ is square and   unknown. The only information available for control
 	design is that $(\mathbf{B}+ \mathbf{B}^{\top})$ is  positive definite and symmetric.
 \end{assumption}
 \begin{theorem}\label{theorem 5}
 	Under Assumptions \ref{assumption 1} and \ref{assumption7}, the closed-loop system consisting of (\ref{dX=BU+F}) and the control law (\ref{u-square-MIMO}) is PT stable in the sense of Definition \ref{definition4} and  all internal signals are bounded over the time interval $[0,T)$, 
 	\begin{equation}\label{u-square-MIMO}
 	\begin{aligned}
 	U=-kZ-\theta Z  \|\Phi\|^2
 	\end{aligned}
 	\end{equation}
 	where $k>0$, $\theta>0$, $Z=\frac{TX}{T-t} $, and $\Phi= \Psi+\dot{\mu}\mu^{-2}\|Z\| $.
 \end{theorem}
 \textit{Proof:}  
 Consider  $V=\frac{1}{2w_B}Z^{\top}Z$ with $w_B$ being an unknown positive constant, then 
 \begin{equation}\label{dV-MIMO-1}
 \begin{aligned}
 \dot V=&\frac{1}{w_B}\mu Z^{\top}   \left(\mathbf{B}U+F+\dot{\mu}\mu^{-2}Z\right)\\
 =&\mu Z^{\top} \left(\frac{\mathbf{B}+\mathbf{B}^{\top}}{2w_B}-\frac{\mathbf{B}-\mathbf{B}^{\top}}{2w_B}\right) \left(-kZ-\theta Z\| \Phi \|^2\right)\\
 &+      \frac{1}{w_B}\mu Z^{\top}\left(F+\dot{\mu}\mu^{-2}Z\right).
 \end{aligned}
 \end{equation}
 In light of Assumption \ref{assumption7}, there exists some unknown constant $w_B>0$, such that $0<w_B\leq\frac{1}{2}\lambda_{\min} (\mathbf{B}+\mathbf{B}^{\top}) $. 
 Therefore,  
 \begin{equation}\label{MIMO-eig}
 \frac{1}{2w_B}Z^{\top}(\mathbf{B}+\mathbf{B}^{\top})Z\geq \|Z\|^2.
 \end{equation}
 In addition, $(\mathbf{B}- \mathbf{B}^{\top})$ is skew symmetric and hence $Z^{\top}(\mathbf{B}-\mathbf{B}^{\top})Z=0,~\forall Z\in\mathbb{R}^{n}$.  Now, it follows from (\ref{dV-MIMO-1}) that
 \begin{equation}\label{dV-MIMO-2}
 \begin{aligned}
 \dot V 
 \leq&-k \mu \|Z\|^2-\theta \|Z\|^2\| \Phi \|^2 +      \frac{1}{w_B}\mu Z^{\top}\left(F+\dot{\mu}\mu^{-2}Z\right).
 \end{aligned}
 \end{equation}
With Young's inequality, we get $\frac{1}{w_B}\mu Z^{\top}(F+\dot\mu\mu^{-2}Z)\leq\frac{1}{w_B}\mu \|Z\|(d(t)\Psi+\dot\mu\mu^{-2}\|Z\|) \leq \mu \|Z\| \Delta \Phi \leq \theta \mu \|Z\|^2\|\Phi\|^2+\frac{\mu\Delta^2}{4\theta}$, with $\theta>0$, $\Delta=\frac{1}{w_B}\times\max\{1,\sup\{|d(t)|\}\}$ and $\Phi =\dot\mu\mu^{-2}\|Z\|+\Psi$. Therefore, we have 
 \begin{equation}
 \dot V\leq -k\mu \|Z\|^2+\frac{\mu\Delta^2}{4\theta}=-2w_Bk\mu(t) V+\frac{\mu\Delta^2}{4\theta}.
 \end{equation} 
 It follows from Proposition \ref{lemma1} that $V\in\mathcal{L}_{\infty}[0,T)$. Using the analysis similar to that below (\ref{dV=-kmuV+mud}), one can conclude that all signals are bounded over $[0,T)$ and $X(t)\rightarrow \mathbf{0}$  as $t\rightarrow T$. Therefore, (\ref{dX=BU+F}) is PT stable in the sense of Definition \ref{definition3}. $\hfill\blacksquare$
 
 \subsection{Non-Square system}
 Now we consider a non-square MIMO system $\dot{X}=\mathbf{B}U+F$ satisfying the following Assumption:
 \begin{assumption}[{[\citen{2022-ye-tac}]}]\label{assumption8}
The high frequency gain matrix $\mathbf{B}(X,t)\in\mathbb{R}^{n\times m}$ can be characterized as $\mathbf{B}(X,t)=\mathbf{A}(X,t)\mathbf{M}(X,t)$, where $\mathbf{M}\in \mathbb{R}^{m\times m}$ is uncertain yet possibly  asymmetric  and  $\mathbf{A}\in\mathbb{R}^{n\times m}$ is a known matrix with full row rank. The message usable for synthesis is that $\mathbf{A}(\mathbf{M}+\mathbf{M}^{\top})\mathbf{A}^{\top}$ is symmetric and positive definite.
 \end{assumption}
 
 Under Assumption \ref{assumption8}, we get a new MIMO system as follows 
 \begin{equation}
 \dot{X}=\mathbf{A}\mathbf{M} U+F(X,t)
 \end{equation}
 where  $U\in\mathbb{R}^m$ and $X\in \mathbb{R}^n$  are the input and the state vector, respectively.
 
 According to Assumption \ref{assumption7}, we known that   the positive definiteness of $(\mathbf{B}+\mathbf{B}^{\top})$ ensures that $\lambda_{\min}(\mathbf{B}+\mathbf{B}^{\top})$ is always positive and there exists some positive unknown constant $w_A$, such that
 $0<w_A\leq \frac{1}{\|\mathbf{A}\|}\lambda_{\min} (\mathbf{A}(\mathbf{M}+\mathbf{M}^{\top})\mathbf{A}^{\top}) $.
 \begin{theorem}\label{theorem 6}
 	Under Assumptions \ref{assumption 1} and \ref{assumption8}, the closed-loop system consisting of (\ref{dX=BU+F}) and the control law (\ref{u-nonsquare-MIMO}) is PT stable in the sense of Definition \ref{definition4} and  all internal signals are bounded over  $[0,T)$,  
 	\begin{equation}\label{u-nonsquare-MIMO}
 	\begin{aligned}
 	U=-\frac{\mathbf{A}^{\top}}{\|\mathbf{A}\|}\left(kZ+\theta Z  \|\Phi\|^2\right)
 	\end{aligned}
 	\end{equation}
 	where $k>0$, $\theta>0$, $Z=\mu X$ with $\mu (t)=\frac{T}{T-t}$ and $\Phi= \Psi+\dot{\mu}\mu^{-2}\|Z\| $.
 \end{theorem}
 \textit{Proof:}  This proof is omitted as it is straightforward by taking  the analysis in  the proof of Theorem \ref{theorem 5}.  The difference is that we need to replace the inequality $Z^{\top}\frac{\mathbf{B}+\mathbf{B}^{\top}}{2}Z\geq w_B\|Z\|^2$ in (\ref{MIMO-eig}) with $Z^{\top}\frac{\mathbf{A}(\mathbf{M}+\mathbf{M}^{\top})\mathbf{A}^{\top}}{2\|\mathbf{A}\|}Z\geq w_A\|Z\|^2$. $\hfill\blacksquare$ 
 
 \begin{remark}
 The main challenges in designing a PT controller for a high order MIMO system are how to cope with the unknown nonlinear perturbations due to the unknown control matrix and how to  relax the assumptions on the control matrix in order to make more general control algorithms.
 \end{remark}

 \section{latest developments in Prescribed-time   control}\label{section5}
 In this section, we aim to present a  literature survey of the foundations of PT decentralized control theory. Knowledge of graph theory can be found in any of the papers about multi-agents, which we have omitted here due to space constraints.
 
 The idea of using time-varying feedback to obtain PT stability has already appeared in early distributed control and has accomplished a large diffusion in recent years. For example, PT consensus on single and double integrator dynamics cases \cite{2012-xieguangmin-PTMAS,2019-jingangshan-PTMAS}; 
PT consensus under undirected/directed graph and PT containment under multiple leaders of first-order networked multi-agent systems\cite{2019-cyber-wangyujuan-Prescribed};  
 leader-following control of high-order multi-agent systems\cite{2018-wangyujuan-Leader}; 
PT consensus via time base generator\cite{2018-TGB-MAS};
 cluster synchronization of complex networks\cite{2020-cyber-pcy};
 lag consensus of second order leader-following  multi-agents\cite{2021-ren-prescribed}; 
PT consensus observer for high-order multi-agents\cite{2021-gongxin-prescribed-observer};
PT bipartite consensus tracking\cite{2020-hanqinglong-PTMAS,2021-gongxin-ptc-bipartite};  
PT consensus over time-varying graph via time scaling\cite{2017-kanzhen-finite}, and then generalized in  \cite{2018-yucelen-finite,2021-arabi-finite,2021-tran-finite,2021-kurtoglu-time}, in which, PT formation tracking, leader-following control, uncertain multi-agent dynamics, multi-agent rigid body system, are considered.

 \subsection{Prescribed-time consensus protocol}
 Consider a multi-agent system where the dynamics of each sub-agent is a single integrator:
 \begin{equation}\label{system-MAS}
 \dot{x}_{i}=u_{i}, ~~~i=1,2, \cdots, n.
 \end{equation}
 The general consensus protocol is:
 \begin{equation}\label{protocol}
 u_{i}=-k\sum_{j=1}^{n} a_{i j} \operatorname{sgn}(x_j-x_i)  |x_{j}-x_{i}|^{\alpha_{ij}}, \quad 1 \leq i \leq n 
 \end{equation}
 where $0 \leq \alpha_{i j}\leq1$ and $k>0$. Obviously, protocol (\ref{protocol}) covers several  common cases:
 \begin{itemize}
 	\item when $\alpha_{i j}=1$, it simplifies  to the classical asymptotic consensus protocol studied in \cite{2004-olfati-consensus}; then the original system can be abbreviated as $\dot{x}=-\mathbf{L} x$, where $x=[x_1,x_2,\cdots,x_n]^{\top}\in\mathbb{R}^n$ and $\mathbf{L}$ is the Laplace matrix of the system, and $\mathbf{L}=\left[l_{i k}\right]_{n \times n}$,
 	\begin{equation}
 	l_{i k}=\left\{\begin{array}{cc}
 	\sum_{j \in \mathcal{N}_{i}} a_{i j}    , & k=i \\
 	-a_{i k}  , & k \neq i.
 	\end{array}\right.
 	\end{equation}
 	 Meanwhile,  the Laplace matrix $\mathbf{L}$ has only one zero eigenvalue and all other eigenvalues with positive real parts if and only if the corresponding directed graph $\mathcal{G}$ contains a spanning tree \cite{2007-renwei}. 	
 	\item  when $\alpha_{i j}=0$, it  corresponds to the discontinuous FT consensus protocol outlined in \cite{2011-chen-finite}.
 	\item when $0<\alpha_{i j}<1$, it reduces to the continuous however nonsmooth FT consensus protocol established in \cite{2009-xiao-finite}. 
 \end{itemize}
 It is important to note that with $0<\alpha_{i j}<1$, the finite settling time $T$  is determined by Proposition \ref{proposition1} as  $T=  \frac{V^{1-\alpha}(0)}{c(1-\alpha)}$
 with $c>0$ being some constant  associated with  the design parameters $k$, $\alpha_{i j}$, and $\lambda_{2}(\mathbf{L})$\footnote{$\lambda_{i}(\mathbf{L})$ denotes the $i$-th minimum eigenvalue  of	the Laplace matrix $\mathbf{L}$.} (which relies on the  structure of $\mathcal{G}$).   There are several issues associated with the settling time $T$:
 \begin{itemize}
 	\item The settling time $T$ is affected by  design parameters $k$ and $\alpha_{ij}$, the initial state $V(0)$, as well as the topological structure.
 	\item To produce a lower $T$, one can increase $k$ or decrease $\alpha_{ij}$ (creating a larger $c$ or a smaller $\alpha$), but the control effort increases with a smaller   $\alpha_{ij}$.  
 	\item If a settling time $T$ is imposed, it is necessary to try to find the relevant parameters $c$ and $\alpha$ based upon $V(0)$ from Proposition \ref{proposition1}, which cannot be explicitly pre-set because $\alpha_{ij}$ is implicitly involved in the function and the initial condition may be unknown.
 \end{itemize}
 The following PT consensus protocol circumvents all the aforementioned shortcomings\cite{2019-cyber-wangyujuan-Prescribed}:
 \begin{equation}\label{MAS-u}
 u_i= -\left(k+c\frac{\dot \mu}{\mu}\right)e_i,~~~i=1,2,\cdots,n
 \end{equation}
 where $k>0$, $c$ is a  parameter will be designed later, $e_i=\sum_{j \in \mathcal{N}_{i}}a_{ij}(x_i-x_j)$ is the local neighborhood error, and $\mu(t)$ is well defined on $[0,T)$ as in (\ref{state-scaling}).
 \begin{theorem}\label{theorem7}
 	Consider system (\ref{system-MAS}) in conjunction with the protocol (\ref{MAS-u}). If the the  graph $\mathcal{G}$ is undirected and connected, and the   design parameter $c$ is selected as $c\geq 1/\lambda_{2}(\mathbf{L})$, then the consensus is attained in prescribed-time, namely   
 	\begin{equation}
 	\|\chi(t)\|\leq\frac{1}{\mu(t)}\|\delta(0)\|e^{-k\lambda_{2}(\mathbf{L})t},~~\forall t\in[0,T)
 	\end{equation} 
 	where $\chi=[\chi_1,\chi_2,\cdots,\chi_n]\in{\mathbb{R}^n}$ and $\chi_i=x_i-\frac{1}{n}\sum_{i=1}^{n}x_i$. Furthermore, the control input remain $\mathcal{C}^1$ smooth and bounded over $[0,T).$
 \end{theorem}
 \subsection{Prescribed-time containment protocol}
 When the communication topology structure involves multiple leaders, the  containment  control  can be naturally evolved from the consensus control. In this subsection, the achieved consensus result is extended to the scenario of containment.
 \begin{theorem}\label{theorem8}
 	Consider system (\ref{system-MAS}) in conjunction with  the protocol (\ref{MAS-u}). If the  graph $\mathcal{G}$ has a directed spanning tree leaded by the root node $x_i$, and the design parameter $c$ is selected as $c\geq 2\lambda_{\max}(\tilde{\mathbf{P}})/\lambda_{1}(\tilde{\mathbf{Q}})$,
 	then, for $\forall t\in[0,T)$, the containment is attained in  prescribed-time, namely 
 	\begin{equation*}
 {\small 	\|\tilde{Z}(t)\|\leq\frac{1}{\mu(t)}\sqrt{\frac{\lambda_{\max}(\tilde{\mathbf{P}})}{\lambda_{\min}(\tilde{\mathbf{P}})}}\left\|\mathbf{L}_1^{-1}\otimes \mathbf{I}_m\right\|\|\tilde{E}(0)\|e^{\frac{-k\lambda_{1}(\tilde{\mathbf{Q}})t}{2\lambda_{\max}(\tilde{\mathbf{P}})}}}
 	\end{equation*} 
 	where $\mathbf{L}_1\in\mathbb{R}^{(n-1)\times(n-1)}$ is a nonsingular matrix with all  eigenvalues satisfying $\lambda_{i}(\mathbf{L}_1)>0,~i=1,\cdots,n-1$, whose specific expression can be obtained according to the Laplace matrix $\mathbf{L}$, namely 
 	\begin{equation*}
 	{\small \mathbf{L}=\left[\begin{array}{cc}
 	0 & \mathbf{0}_{1 \times(n-1)} \\
 	\mathbf{L}_{2} &\mathbf{L}_{1}
 	\end{array}\right],}
 	\end{equation*}
 	and $\tilde{Z}=[z_2,z_3,\cdots,z_n]^{\top}\in\mathbb{R}^{n-1}$ with $z_i=x_i-x_1$, $\tilde{E}=[e_2,e_3,\cdots,e_n]^{\top}\in\mathbb{R}^{n-1}$ and $\tilde{\mathbf{Q}}=\tilde{\mathbf{P}}\mathbf{L}_1+\mathbf{L}_1^{\top}\tilde{\mathbf{P}}$  with $\tilde{\mathbf{P}}=\operatorname{diag}\{p_2,\cdots,p_n\}$ and  $[p_2,\cdots,p_n]^{\top}=(\mathbf{L}_1^{\top})^{-1}\mathbf{1}_{n-1}$. Furthermore, the control $u_i$ remains $\mathcal{C}^1$ smooth and bounded over $[0,T).$
 \end{theorem}
 \textit{Proof:} The proofs of Theorems \ref{theorem7}-\ref{theorem8} are omitted as they can be  found in \cite{2019-cyber-wangyujuan-Prescribed}. $\hfill\blacksquare$

 \section{Comparison and conclusion}\label{section6}
 To close  this article, we recap the connection between the PT results to FT results, and discuss the related numerical implementation issues. In addition, we compare the differences between typical FT/FxT/PdT and PT controllers via simulation on a double integrator. Finally, we conclude by giving some future research challenges. 
 \subsection{Controller structure}
 Consider the first-order integrator as follows:
 \begin{equation}\label{dx=u}
 \dot x=u   ,~~~x(0)=x_0
 \end{equation}
From \cite{2020-zhoubin-prescribed}, one can immediately obtain a time-varying feedback based PT controller as
 \begin{equation}\label{u_p}
 u_{\operatorname{prescribed}}=-\frac{kx}{T-t} ,~k\geq 1.
 \end{equation}
 Also, from \cite{1998-bhat}, we get  the classical FT autonomous controller as
 \begin{equation}\label{u_f}
 u_{\operatorname{finite}}=-k\operatorname{sgn}(x)|x|^{\alpha},~0<\alpha<1,~ k>0,
 \end{equation}
 then the solution of (\ref{dx=u}) with (\ref{u_f}) is 
 \begin{equation*}
\left\{
\begin{array}{l}
 x(t)= 
 \left(|x_0|^{1-\alpha}-k(1-\alpha)t\right)^{\frac{1}{1-\alpha}}\operatorname{sgn}(x_0), ~ t\in[0,T_f)\\ 
 x(t)=0,~~~~~~~~~~~~~~~~~~~~   ~~~~~~~~~~~~~~~~~~~~~~t\in [T_f,+\infty)
\end{array}
\right.
 \end{equation*}
 with $$T_f=\frac{1}{k}\frac{1}{(1-\alpha)}|x_0|^{1-\alpha}.$$ Therefore, we have $|x_0|^{1-\alpha}=kT_f(1-\alpha)$, and the control law (\ref{u_f}) can be rewritten as 
 \begin{equation}\label{25}
 \begin{aligned}
 u_{\operatorname{finite}}=&-k\operatorname{sgn}(x)|x|^{\alpha-1}|x|\\
 =&-k\left|\left(|x_0|^{1-\alpha}-k(1-\alpha)t\right)^{\frac{1}{1-\alpha}}\right|^{\alpha-1}x\\
 =&-\frac{x}{(1-\alpha)(T_f-t)}.
 \end{aligned}
 \end{equation}
 Note that if we choose $k\equiv \frac{1}{1-\alpha}$, the PT controller (\ref{u_p}) becomes the FT controller (\ref{u_f}), which means that the FT controller  (\ref{u_f})  is indeed a special case of the PT controller (\ref{u_p}). In fact, they share the same property that the control gain tends to $\infty$ as $t\rightarrow T$.  As a matter of fact,  all FT controllers (including FxT controllers, PT controllers, and PdT controllers) share this property. Also note that the magnitude of the PT control input (\ref{u_p}) (consisting of a high-gain function $-\frac{k}{T-t}$ and a feedback signal $x$) does not become large when the feedback signal decays faster than the high-gain function grows.

\subsection{Discussion on Implementation}
In the implementation of FT control algorithms, it is necessary to introduce sign function $\operatorname{sgn}(x)$  to avoid singularity when $x(t)=0$.  For example, the control law $u=-x^{1/3}$  is programmed to be replaced by  $u=-\operatorname{sgn}(x)|x|^{1/3}$. 
Two effective ways of implementing PTC are:
 \begin{itemize}
 	\item Letting  $T=T_{\operatorname{s}}$ (scheduled time) $+\epsilon$ (small constant)  so that the controller works for the scheduled time;
 	\item Setting an upper bound on the scaling function $\mu(t)$ before  the time variable approaches the desired preset time $T$.
 \end{itemize} 
 Anyhow, unbounded control gain will not cause unbounded control input, and many simulation results show that the PT regulation is achieved with a suitable control effort, without  an exorbitant price. Both of the above implement methods slightly sacrifice the control precision in favor of promoting practical implementation by avoiding unbounded gains.
The major concern for time-varying feedback control is its robustness against measurement noise.

 To show the characteristics and the differences of FT, FxT, PdT and PT control schemes, we consider a double integrator for numerical simulation. The system model is a double integrator as follows:
 \begin{equation}
 \dot{x}_1=x_2,~~~\dot{x}_2=u.
 \end{equation}
 The FT controller [\citen{1998-bhat}, Example 1], the FxT controller [\citen{2020Polyakov}, Example 5.11], the PdT controller  [\citen{2018-Sanchez-predefined-time}, Example 4.2], and the PT controller  \cite{2021-TAC-huachanghun-prescribed-adaptive} are shown below:  
 {\small \begin{itemize}
 	\item $u_{\operatorname{finite}}=-x_2^{\frac{1}{3}}-\big(x_1+\frac{3}{5}x_2^{\frac{5}{3}}\big)^{\frac{1}{5}}$;
 	\item $u_{\operatorname{fixed}} = -\frac{1+3x_1^2}{2}\operatorname{sgn}(s)-(s+s^3)^{[\frac{1}{2}]}  
 	$ with $	s=x_2+(x_2^{[2]}+x_1+x_1^3)^{[\frac{1}{2}]}$ and $v^{[a]}=|v|^{a}\operatorname{sgn}(v)$;
 	\item $u_{\operatorname{predifined}}=-\frac{\partial \Phi_1(x_1,T_1)}{\partial x_1}x_2-\Phi_2(\sigma,T_2)$ with $\sigma= \Phi_1 +x_2$ and $\Phi_i(v,T_i)=\frac{5}{2T_i}\exp({|v|^{\frac{2}{5}}}){|v|^{\frac{3}{5}}}\operatorname{sgn}(v)$;
 	\item  
 	$u_{\operatorname{prescribed}}=-2\mu(x_2+3\mu x_1)-3\mu^2 x_1-3 \mu x_2$ for $t\in[0,T)$ and $u_{\operatorname{ps}}=0$ for $t\in[0,\infty)$ with $\mu=1/(T-t) $.
 \end{itemize}}
Two scenarios are considered for simulation: $x_1(0)=0.2$ and $0.4$ and $x_2(0)=-0.2$ and $0$. Figs. 3-6 illustrate the simulation results, from which, it can be seen that the PT controller achieves FT regulation in $T = 1s$, whereas the settling time of the FT controller depends on the initial conditions, the FxT controller depends on the design parameters and the upper bound on the settling time is overestimated, and the PdT controller requires a larger control input when the initial conditions become slightly larger, and a slighter overestimation of the settling time can be observed. Besides, from Figs. 3-6, it is observed that the PT controller exhibits smoother control action, avoiding the  chatting phenomenon as reflected in Figs. 3-5. These simulation results show to some extent the superiority of PT control compared to the other three control methods.
 \begin{figure}[!h]
 	\begin{center}
 		\includegraphics[height=2.7cm]{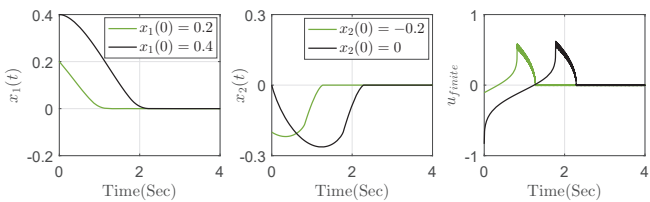}
 		\caption{FT control for a double integrator.}
 	\end{center}
 \end{figure}
\vspace{-0.7cm} 
\begin{figure}[!h]
 	\begin{center}
 		\includegraphics[height=2.7cm]{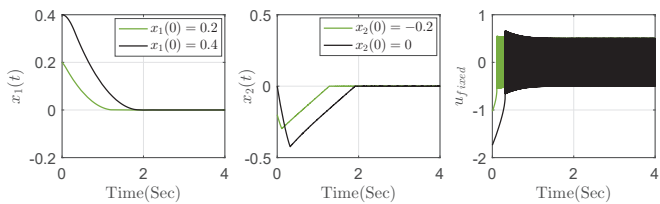}
 		\caption{{FxT control for a double integrator with the   settling time $t\leq T_{\max}= {\pi} +\frac{\pi}{\sqrt{2}} \approx 5.363s $.}}
 	\end{center}
 \end{figure} 
 \vspace{-0.7cm}
 \begin{figure}[!h]
 	\begin{center}
 		\includegraphics[height=2.8cm]{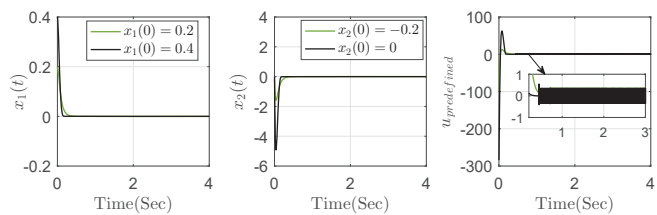}
 		\caption{{PdT control for a double integrator with the  settling time $t\leq T_{\max}=T_1+T_2=(0.2+0.8)s=1s $.}}
 	\end{center}
 \end{figure} 
\vspace{-0.7cm}
 \begin{figure}[!htbp]
 	\begin{center}
 		\includegraphics[height=2.6cm]{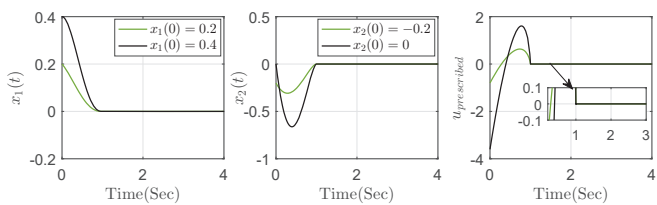}
 		\caption{{PT control for a double integrator with the settling time $ T = 1s$.}}
 	\end{center}
\end{figure}

 \subsection{Challenges and Future Opportunities}\label{section7}
 The core idea in PT control is to design a suitable time-varying gain that goes to $\infty$ as the system converges to zero, which is derived from proportional navigation in strategic and tactical missile guidance applications. In PT control, we require more than that, not only to regulate the state precisely to zero but also to ensure that the control signal is bounded, while completely rejecting external disturbance. 
 The potential topics on PT control for complex dynamic plants include (but not limited to):
 \begin{itemize}
 	\item  Although adaptive PT control is investigated in \cite{2021-TAC-huachanghun-prescribed-adaptive}, the system unknown parameter must remain unchanged , while the control coefficient must be accessible. In some modern practical applications (such as high-performance robots\cite{sci-bird,petersen,sci-walk-robot}), such assumptions may not be satisfied since we know that systems with changing structures usually have time-varying system parameters and that the inertia matrix (which can be viewed as the control influence gain) of a robotic system is usually unknown. It is necessary and challenging to develop more powerful solutions to meet such scenarios. 
 	\item Output feedback schemes often imply low cost, which is very attractive in practical applications (especially for large-scaled/networked/multi-agent systems). However, the existing results can only achieve  output feedback PT stabilizing control for some special systems (e.g., linear systems),   it is therefore important to explore the output feedback based PT control for more general systems. 
 	\item In addition to the PT stabilizing control, the study of the PT tracking control is more general, however, when tracking is considered, the desired trajectory to be tracked would give rise to extra time variation and/or uncertainties and hence brings technical obstacles. How to improve the PT control algorithm so that it completely rejects these non-vanishing uncertainties (which may be generated by the desired tracking signal, may come from some external noise, or may be inherent in the physical model) is an interesting future research topic.
 	\item As for PT control for multi-agent systems, it  is  interesting to generalize the simple framework on first or second-order integrators to agents having high-order uncertain nonlinear dynamics and to investigate PT decentralized control algorithms under complex communication topologies, as well as to study how to achieve consensus with as little information interaction between agents as possible without losing controllability.
 	\item The study of more types of system models, more low-conservative control algorithms or the pursuing for better control performance of closed-loop systems are all interesting future research topics in the field of PT control. 
 \end{itemize}


\end{document}